\documentclass[prx,aps,twocolumn,nopacs,superscriptaddress,longbibliography,nofootinbib]{revtex4-1}

\usepackage{amsmath}  \usepackage{amssymb}  \usepackage{amsfonts}  \usepackage{bm}  \usepackage{bbm}   \usepackage{bbold}   \usepackage{braket}  \usepackage{color}  \usepackage{comment}  \usepackage{dcolumn}  \usepackage{enumerate}  \usepackage{epsfig}  \usepackage{gensymb}  \usepackage{graphicx}  \usepackage{indentfirst}  \usepackage{lmodern}  \usepackage{mathrsfs}  \usepackage{mathtools}  \usepackage{psfrag}  \usepackage{pst-all}  \usepackage{soul}  \usepackage{xcolor}
\usepackage{float}
\usepackage[colorlinks,linkcolor=blue,citecolor=blue,urlcolor=blue,hyperindex,driverfallback=dvipdfm]{hyperref}  \usepackage[T1]{fontenc}

\def\ii{i}  \def\ee{e}
  
\def\rb{{\bf r}}  \def\Rb{{\bf R}}  \def\ub{{\bf u}}  \def\vb{{\bf v}}
    \def\zz{\hat{\bf z}}  \def\nn{{\hat{\bf n}}}  \def\rr{{\hat{\bf r}}}
\def\nn{\hat{\bf n}}  
  
  \def\kpar{k_\parallel}  
\def\Qb{{\bf Q}}       \def\Db{{\bf D}} \def\db{{\bf d}}
  \def\fb{{\bf f}}      \def\qb{{\bf q}}   \def\sb{{\bf s}}    
\def\me{m_e}  
\def\Eb{{\bf E}}      \def\Ab{{\bf A}}
  \def\jb{{\bf j}}
  
    \def\EF{{E_F}}
      
\def\wp{\omega_p}
  
\def\nn{{\hat{\bf n}}}  \def\wb{{\bf w}}
\def\Cs{C_s}  \def\Cp{C_p}    \def\qb{{\bf q}}   \def\qmk{{Q_{m k}}}

\begin{document}
\title{Toward Optimum Coupling between Free Electrons and Confined Optical Modes}

\author{Valerio~Di~Giulio}
\affiliation{ICFO-Institut de Ciencies Fotoniques, The Barcelona Institute of Science and Technology, 08860 Castelldefels (Barcelona), Spain}
\author{Evelijn Akerboom}
\affiliation{Center for Nanophotonics, NWO-Institute AMOLF, Science Park 104, 1098 XG Amsterdam, The Netherlands}
\author{Albert Polman}
\affiliation{Center for Nanophotonics, NWO-Institute AMOLF, Science Park 104, 1098 XG Amsterdam, The Netherlands}
\author{F.~Javier~Garc\'{\i}a~de~Abajo}
\email[Corresponding author: ]{javier.garciadeabajo@nanophotonics.es}
\affiliation{ICFO-Institut de Ciencies Fotoniques, The Barcelona Institute of Science and Technology, 08860 Castelldefels (Barcelona), Spain}
\affiliation{ICREA-Instituci\'o Catalana de Recerca i Estudis Avan\c{c}ats, Passeig Llu\'{\i}s Companys 23, 08010 Barcelona, Spain}

\begin{abstract}
Free electrons are unique tools to probe and manipulate nanoscale optical fields with emerging applications in ultrafast spectromicroscopy and quantum metrology. However, advances in this field are hindered by the small probability associated with the excitation of single optical modes by individual free electrons. Here, we theoretically investigate the scaling properties of the electron-driven excitation probability for a wide variety of optical modes including plasmons in metallic nanostructures and Mie resonances in dielectric cavities, spanning a broad spectral range that extends from the ultraviolet to the infrared. The highest probabilities for the direct generation of three-dimensionally confined modes are observed at low electron and mode energies in small structures, with order-unity ($\sim100$\%) coupling demanding the use of $<100$~eV electrons interacting with $<1$~eV polaritons confined down to tens of nm. Electronic transitions in artificial atoms also emerge as practical systems to realize strong coupling to few-eV free electrons. In contrast, conventional dielectric cavities reach a maximum probability in the few-percent range. In addition, we show that waveguide modes can be generated with higher-than-unity efficiency by phase-matched interaction with grazing electrons, suggesting an efficient method to create multiple excitations of a localized optical mode by an individual electron through funneling the so-generated propagating photons into a confining cavity---an alternative approach to direct electron--cavity interaction. Our work provides a roadmap to optimize electron--photon coupling with potential applications in electron spectromicroscopy as well as nonlinear and quantum optics at the nanoscale.
\end{abstract}
\maketitle
\tableofcontents

\section{Introduction}

Probing confined optical excitations with nanometer resolution is important for understanding their microscopic dynamics and properties down to the atomic scale. Currently, energy electron-loss spectroscopy (EELS) and cathodoluminescence (CL) spectroscopy permit studying photonic nanostructures with an unparalleled combination of spatial and energy resolution \cite{KLD14,LTH17,HNY18,LB18,HKR19,HHP19,HRK20,YLG21}. While the direct emission or absorption of photons by free electrons is forbidden in free space due to energy-momentum conservation, the near-field associated with confined optical modes provides the necessary momentum to break such a mismatch, materializing an inelastic interaction. In EELS, the electron beam (e-beam) couples to both dark and bright optical modes \cite{paper149}, resulting in a loss of energy that is recorded using an electron spectrometer. In CL, one measures the light emission produced by excited bright modes. However, when the spectral features associated with optical modes in the EELS and CL signals are integrated over photon frequency,  one generally finds small excitation probabilities, several orders of magnitude smaller than unity \cite{NWM11,CKS16,WLC18}. This weak-coupling regime is advantageous to resolving clean linear spectra, but it represents a strong limitation when envisioning applications that require multiple excitations, which are needed to trigger a nonlinear response and enter the terrain of quantum optics at the nanoscale.

A practical way of increasing the probability of interaction per electron is exploited in photon-induced near-field electron microscopy \cite{BFZ09,PLZ10,paper151,FES15,HBL16} (PINEM)---a form of stimulated inelastic electron–light scattering (SIELS) in which strong external light fields are aimed at a specimen in coincidence with the electron arrival. While EELS and CL rely on spontaneous excitation processes of previously unexcited optical modes, PINEM (and SIELS in general) is dominated by stimulated transitions of modes that are highly populated through external illumination, generally resulting in a symmetric set of loss and gain sidebands in the electron spectrum, with a probability increased by the mode population relative to EELS and CL \cite{paper114}. Importantly, when employing external laser light, the sidebands form a coherent superposition (i.e., a single electron wave function consisting of an energy comb) evolving with different energy-dependent velocities and eventually resulting in a temporal comb of strongly compressed pulses \cite{FES15,YFR21,paper415}. SIELS thus enables a wide degree of control over the free-electron wave function \cite{KSE16,EFS16,PRY17,MB18,paper332,paper360,FYS20,KLS20}. Remarkably, SIELS holds the promise of achieving a combined attosecond/sub-nanometer resolution in time/space \cite{paper415}. In addition, electron energy-gain spectroscopy \cite{H99,paper114} (EEGS) capitalizes on SIELS to combine the spatial resolution of free electrons with the energy resolution inherited from the spectral precision with which the laser frequency can be tuned. EEGS has been recently demonstrated to map microring \cite{HRF21} and Mie \cite{paper419} optical cavities with $\mu$eV energy resolution. However, SIELS involves the presence of intense optical fields, and therefore, it does not help us to explore nonlinear and quantum phenomena at the few-photon level.

Several theoretical works have explored the use of free electrons to herald the production of single photons \cite{paper180}, entangled photons \cite{paper428}, and other nonclassical light states \cite{paper402,HEK23,DBG23,KF23}, while a recent experiment has demonstrated Fock photon-state generation in optical resonators correlated with energy losses of the electron \cite{FHA22}. A large electron--photon coupling is then required at the single-electron/single-photon level, which is currently achieved in such resonators by matching the phases of the electron excitation current and the light propagating inside a curved waveguide over an interaction distance of several microns \cite{FHA22}. This is the so-called phase-matching approach \cite{paper180,K19,HRF21,FHA22,DDB23}, which provides an alternative method to couple single electrons to confined modes, as we discuss below. Fundamental upper bounds for the maximum electron--mode coupling have been formulated for these and other geometries \cite{YMR18,XCL24,Z24}.

In this article, we determine the conditions under which unity-order coupling is possible between free electrons and optical modes confined in all three spatial directions. We find that low-energy modes (sub-eV) confined down to small structures (tens of nm) can be excited with high probability ($\sim100\%$) by low-energy electrons (tens of eV). Suitable systems to support the required modes are two-dimensional-material nanostructures made of polaritonic films (e.g., graphene), while electronic transitions in artificial atoms also emerge as a plausible platform. We illustrate our results with a comprehensive exploration of the scaling properties of EELS and CL probabilities as a function of the size of the polariton-supporting structure and the electron velocity, translated into simple analytical expressions that should facilitate the task of designing optimum electron--cavity configurations. We further present extensive numerical simulations for a wide range of relevant materials and nanostructure morphologies. This analysis is supplemented by a discussion of free-electron coupling to waveguide modes, which we argue to offer a viable approach to generating multiple excitations in a given three-dimensionally confined mode by an individual electron when the so-generated propagating photons are funneled into a confining optical cavity. Our results support the use of free electrons as pivotal ingredients to manipulate confined optical excitations and provide a roadmap to optimize electron--cavity coupling and reach the sought-after nonlinear regime triggered by an individual electron.

\section{Results and Discussion}

In what follows, we present results based on a classical treatment of the electron as a moving point charge. Commonly, electron spectra are dominated by coupling to bosonic excitations in a specimen, for which quantum theory shows that the associated EELS and CL probabilities follow Poissonian distributions with an average number of excitations that coincides with the classically calculated probabilities \cite{SL1971,paper228,paper339}. For example, for a bosonic mode $i$ of energy $\hbar\omega_i$ initially prepared in the ground state, the mode-integrated EELS probability $P_{{\rm EELS},i}$, which is calculated from classical theory as shown below, must be interpreted as a Poissonian distribution of loss features in the EELS spectra at energies $n\hbar\omega_i$ with probabilities $P_{{\rm EELS},i}^n\ee^{-P_{{\rm EELS},i}}/n!$, where $n=0,\cdots,\infty$ indexes the Fock state $\ket{n}$ resulting from the interaction with the electron. EELS and CL measurements typically reveal small probabilities $\ll1$, so that only the $n=1$ state becomes relevant. Here, we investigate certain conditions under which the classical excitation probability is greater than unity, and therefore, for modes following bosonic statistics, we need to understand such a probability as the average population of a Poissonian distribution.

\subsection{Scaling of the EELS Probability}

One of the central quantities analyzed in this work is the frequency-integrated EELS probability $P_{\rm EELS}$ (i.e., the fraction of electrons that lose some energy after interaction with a specimen). We assume high electron kinetic energies (KEs) compared with the excitation energies so that the electron velocity vector $\vb$ remains approximately constant during the interaction (nonrecoil approximation \cite{paper149}). We further consider an incoming electron wave function $\psi(\rb)$ consisting of a narrow distribution of wave vectors peaked around a central value $\qb_0=\me\vb\gamma/\hbar$, where $\gamma=1/\sqrt{1-\beta^2}$ and $\beta=v/c$. In particular, we are interested in the probability of exciting an isolated resonance mode of energy $\hbar\omega_i$, which, upon integration over the corresponding spectral width (indicated by a subscript $i$ in the integration symbol), is given by \cite{RH1988,paper149}
\begin{align}
P_{{\rm EELS},i}(v)=\int_i d\omega\int d^3\rb\, |\psi(\rb)|^2 \Gamma_{\rm EELS}(\Rb,v,\omega), \label{PpsiG}
\end{align}
where
\begin{align}
\Gamma_{\rm EELS}(\Rb,v,\omega)=&4\alpha c \int_{-\infty}^\infty dz \int_{-\infty}^\infty dz' \, \cos\bigg[\frac{\omega}{v} (z-z')\bigg] \nonumber\\
&\times{\rm Im}\{-G_{zz}(\Rb,z,\Rb,z',\omega)\} \label{eels}
\end{align}
is the frequency-resolved EELS probability \cite{paper149}, $\alpha\approx1/137$ is the fine structure constant, and we set $\vb=v\zz$ without loss of generality. We explicitly indicate the dependence on velocity $v$ because this parameter is playing an important role in the present work. In Eq.~(\ref{eels}), the composition and geometry of the specimen are encapsulated in the electromagnetic Green tensor $G(\rb,\rb',\omega)$, implicitly defined by the equation
\begin{align}
\nabla \times \nabla \times& G(\rb,\rb',\omega)-k^2\epsilon(\rb,\omega) G(\rb,\rb',\omega) \nonumber\\
&=-\frac{1}{c^2}\delta(\rb-\rb'), \label{gt}
\end{align}
where $k=\omega/c$ and we restrict our analysis to materials characterized by a nonmagnetic, isotropic, local permittivity $\epsilon(\rb,\omega)$.

We start our discussion by studying the behavior of the EELS probability when spatially scaling the system by a dimensionless factor $\mu$ ($<1$ for compression and $>1$ for expansion). Specifically, we consider two types of systems:
\begin{itemize}
\item[(\textit{i})] {\it General structures in the quasistatic limit [3D quasistatic].} Three-dimensional (3D) structures made of a material described by a complex, frequency-dependent permittivity $\epsilon(\omega)$ and supporting resonances at optical wavelengths much larger than their characteristic size $D$. For these systems, scaling laws can be formulated in the quasistatic regime (i.e., $\omega D/c\ll1$).
\item[(\textit{ii})] {\it Dielectric cavities [3D dielectrics].} Structures made of a lossless dielectric material that is characterized by a real, frequency-independent permittivity $\epsilon$. Scaling laws can be established for these types of systems with full inclusion of retardation effects.
\end{itemize}

In structures of type (\textit{i}), the dependences on optical frequency and spatial position are decoupled because of the lack of an absolute length scale. Consequently, Eq.~(\ref{gt}) implies that the Green tensor $\tilde{G}$ associated with a system in which all distances are scaled by a factor $\mu$ is related to the Green tensor $G$ of the original system through the equation $\tilde{G}(\rb,\rb',\omega)=\mu^{-3}\,G(\rb/\mu,\rb'/\mu,\omega)$ (see Appendix~\ref{AppendixH} for a self-contained derivation of this result). In addition, Eq.~(\ref{eels}) for the EELS probability can be recast into
\begin{align}
\Gamma_{\rm EELS}(\Rb_0,v,\omega)=&\frac{e^2}{\pi \hbar v^2} \int_{-\infty}^\infty dz\int_{-\infty}^\infty dz' \cos\left[\frac{\omega}{v}(z-z')\right] \nonumber\\
&\times{\rm Im}\{-\mathcal{W}(\Rb_0,z,\Rb_0,z',\omega)\}\label{eelsquasi}
\end{align}
in terms of the screened interaction $\mathcal{W}(\rb,\rb',\omega)$, defined as the electric potential created at $\rb$ by a unit charge placed at $\rb'$ and oscillating with frequency $\omega$. Equation~\ref{eelsquasi} can readily be derived from Eq.~(\ref{eels}) by noticing that the Green tensor can be approximated as $G(\rb,\rb',\omega)\approx \nabla_\rb \otimes \nabla_{\rb'}\mathcal{W}(\rb,\rb',\omega)/4\pi\omega^2$ in this limit (see Appendix~\ref{AppendixH}).

In contrast, for structures of type (\textit{ii}), Eq.~(\ref{gt}) leads to the more general scaling $\tilde{G}(\rb,\rb',\omega)=\mu^{-1}\,G(\rb/\mu,\rb'/\mu,\mu\omega)$. These two scaling laws can be directly applied to the frequency-integrated EELS probability $P_{{\rm EELS},i}(\Rb_0,v)=\int_i d\omega\,[d\Gamma_{\rm EELS}(\Rb_0,v,\omega)/d\omega]$, where the subscript $i$ indicates that we integrate over the spectral range of a given mode $i$, we indicate the lateral position $\Rb_0$ under the assumption of a well-focused electron (i.e., $|\psi(\rb)|^2\approx \delta(\Rb-\Rb_0)$ in Eq.~(\ref{PpsiG})). Indeed, combining the noted properties of the Green tensor with Eq.~(\ref{eels}), we find
\begin{widetext}
\begin{subequations}
\label{Peelsscaling}
\begin{align}
&\tilde{P}_{{\rm EELS},i}(\Rb_0,v)=\frac{1}{\mu}\,P_{{\rm EELS},i}(\Rb_0/\mu,v/\mu) &{\rm [3D~quasistatic]}, \label{Peelsscaling1}\\
&\tilde{P}_{{\rm EELS},i}(\Rb_0,v)=P_{{\rm EELS},i}(\Rb_0/\mu,v) &{\rm [3D~dielectrics]}, \label{Peelsscaling2}
\end{align}
\end{subequations}
where a tilde is used to refer to the scaled system. For dielectrics, Eq.~(\ref{Peelsscaling2}) implies that the electron coupling remains identical in a structure expanded by a factor $\mu>1$ as long as the impact parameter $\Rb_0$ is increased by the same factor. However, when the response of the system depends on frequency, such as in metals and doped semiconductors in the optical regime (see below), the interaction probability can be indefinitely increased by reducing the size of the structure (down to the quasistatic regime), provided we also slow down the electron and bring the trajectory closer to the material.
\end{widetext}

\subsection{Scaling of the CL Probability}

For CL, the process leading to the production of free light far from the specimen can be separated into two steps: coupling of the free electron to an excited state of the structure and subsequent decay to a final state accompanied by the emission of a photon. Under the same assumptions as for EELS, the CL emission probability corresponding to the excitation of a mode $i$ in the specimen can be written as $P_{{\rm CL},i}(\Rb_0,v)= \int_i d\omega \int d\Omega_{\rr}$ $[d\Gamma_{\rm CL}(\Rb_0,v,\omega)/d\omega d\Omega_{\rr}]$, where \cite{paper149}
\begin{align}
\frac{d\Gamma_{\rm CL}(\Rb_0,v,\omega)}{d\omega d\Omega_{\rr}}=\frac{1}{4\pi^2\hbar k} |\fb_{\rr}(\Rb_0,\omega)|^2 \label{cl}
\end{align}
is the angle-resolved probability expressed in terms of the far-field amplitude $\fb_{\rr}(\Rb_0,\omega)=4\pi\ii e\omega \int_{-\infty}^\infty dz'\, G_{\rm ff}(\rr,\Rb_0,z',\omega)\cdot \zz \,\ee^{\ii \omega z'/v}$ and we use the far-field limit of the Green tensor $\lim\limits_{kr\rightarrow \infty} G(\rb,\rb',\omega)=(\ee^{\ii kr}/r) \,G_{\rm ff}(\rr,\rb',\omega)$. For structures of type (\textit{ii}) (dielectric cavities), the absence of absorption in the material implies $P_{{\rm CL},i}=P_{{\rm EELS},i}$ (i.e., all energy losses are converted into radiation) and the scaling expressed in Eq.~(\ref{Peelsscaling2}) applies. In contrast, a general scaling of $P_{{\rm CL},i}$ cannot be obtained for materials described by a frequency-dependent permittivity, not even in the small-particle limit: space and frequency mix due to the finiteness of the speed of light because the radiative part of the CL emission process involves a transverse free photon, which only exists when retardation is taken into consideration. Nevertheless, an interesting relation can still be obtained through a perturbative analysis based on the Dyson equation for the Green tensor by adopting the quasistatic approximation for the response of the structure but incorporating retardation in its outcoupling to radiation (see details in Appendix~\ref{AppendixE}):
\begin{widetext}
\begin{align}
\tilde{P}_{{\rm CL},i}(\Rb_0,v)\approx\mu^2\,P^{(0)}_{{\rm CL},i}(\Rb_0/\mu, v/\mu)\!+\!\mu^3\,P_{{\rm CL},i}^{(1)}(\Rb_0/\mu, v/\mu) +\cdots\quad \quad {\rm [3D~quasistatic]}.\label{Pclscaling}
\end{align}
\end{widetext}
According to Eq.~(\ref{Pclscaling}), the CL emission probability increases when simultaneously increasing the size of the object, the electron velocity, and the impact parameter, in contrast to the behavior of EELS based on Eq.~(\ref{Peelsscaling1}).

\begin{figure*}
\centering
\includegraphics[width=0.85\textwidth]{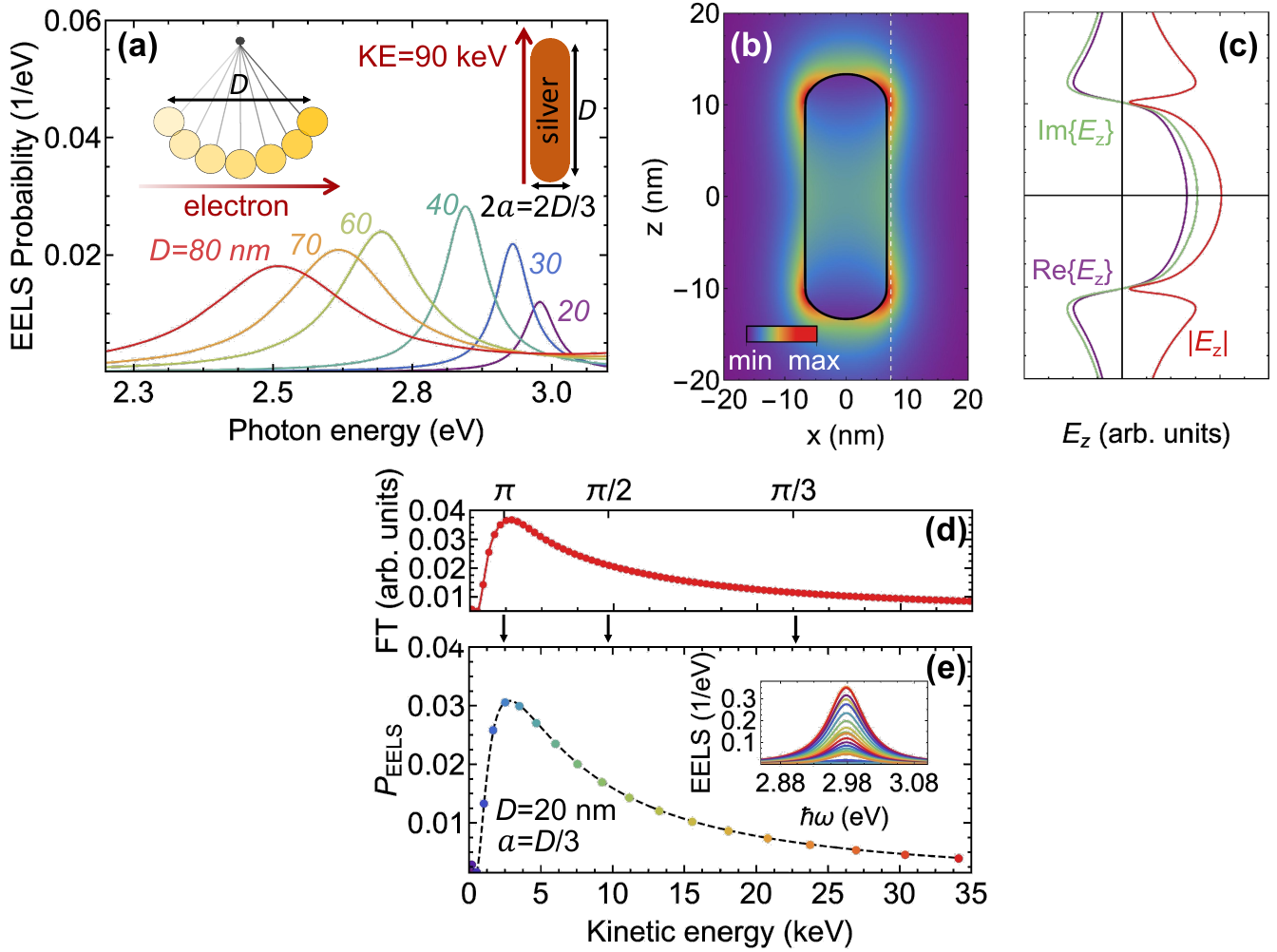}
\caption{{\bf Maximizing plasmon excitation through electron-energy matching.} We consider the interaction of free electrons with the longitudinal dipolar plasmon of silver nanorods. {\bf (a)}~EELS probability experienced by electrons with a kinetic energy (KE) of 90~keV passing parallel and close to the surface of a silver nanorod (see right inset) for several values of the rod length $D$ (see color-coordinated labels) and fixed aspect ratio (length-to-diameter) $D/2a=1.5$. The electron-surface separation is $0.075\,D$. {\bf (b)}~Near-electric-field intensity associated with the longitudinal plasmon mode of the $D=20$~nm rod in (a) (photon energy $\hbar\omega_1=2.97$~eV). {\bf (c)}~Field extracted from (a) along the e-beam direction $z$ (dashed line in (b)). {\bf (d)}~Absolute squared value of the spatial Fourier transform of the field in (c), which depends on KE through the phase $\varphi_1=\omega_1D/v$ (upper horizontal scale), where $v$ is the electron velocity at the KE in the lower horizontal scale. {\bf (e)}~KE dependence of the EELS probability integrated over the color-coordinated spectra shown in the inset under the conditions of (a) for $D=20$~nm.}
\label{Fig1}
\end{figure*}

\subsection{Polaritonic Structures in the Quasistatic Limit}

Structures of type (\textit{i}) can sustain optical resonances when they are made of polariton-supporting materials (e.g., plasmons in metals \cite{S11}, phonon-polaritons \cite{GDV18,MAL18,paper396}, and excitons \cite{LCZ14}). The corresponding mode frequencies $\omega_i$ span a wide spectral range extending from the mid-infrared to the ultraviolet with typical structure sizes $D$ going from a few nm to $\sim1\,\mu$m. We focus on systems operating in the quasistatic limit ($\omega_iD/c\ll1$) and consider electrons traveling without crossing any material boundary. The optical response can then be rigorously expanded in eignmodes $i$ with associated eigenpotentials \cite{OI1989,paper010} $\phi_i(\rb)=\tilde{\phi}_i(\rb/D)/D$, where $\tilde{\phi}_i(\ub)$ are dimensionless, scale-invariant, and material-independent functions of the scaled coordinates $\ub=\rb/D$ (see Eq.~(\ref{phii}) in Appendix~\ref{AppendixA}). Likewise, we define the mode fields $\Eb_i(\rb)=-\nabla\phi_i(\rb)=\tilde{\Eb}_i(\rb/D)/D^2$, also written in terms of dimensionless, scale-invarient vectors $\tilde{\Eb}_i(\ub)=-\nabla_\ub\tilde{\phi}_i(\ub)$. As shown in detail in Appendix~\ref{AppendixD}, the excitation probability of a mode $i$ can be written as
\begin{align}
P_{{\rm EELS},i}(\Rb_0,v)&\approx\frac{\alpha}{\beta} A_i  F_i\left(\varphi_i\right)\quad \quad {\rm [3D~quasistatic]}, \label{quasieelsdep}
\end{align}
where
\begin{subequations}
\label{Fivarphii}
\begin{align}
F_i(\varphi_i)
&=\frac{1}{\varphi_i}\Big|\int_{-\infty}^\infty \,du_z\,\tilde{E}_{i,z}(\Rb_0/D,u_z)\ee^{-\ii \varphi_i u_z}\Big|^2 \label{Fi}
\end{align}
is the squared spatial Fourier transform of the mode electric field along the e-beam direction, taken at a spatial frequency $\omega_i/v$, which permits defining the characteristic phase
\begin{align}
\varphi_i=\omega_i D/v. \label{varphii}
\end{align}
\end{subequations}
In Eq.~(\ref{quasieelsdep}), the material-dependent constant $A_i$ encapsulates the details of the dielectric function and its derivative at the mode frequency $\omega_i$. We note that $A_i$ is independent of size and electron velocity (see Appendix~\ref{AppendixD}). In analogy to the mode fields, the function $F_i(\varphi_i)$ only depends on morphology but not on composition and size. This function plays a fundamental role in determining the electron--mode coupling and its maximum determines the so-called phase-matching condition \cite{ALM23}. In analogy to the optimum overlap between the oscillations of a pendulum (period $2\pi/\omega_i$ and size $D$) and the position of a particle interacting with it (passing with velocity $v$, see Fig.~\ref{Fig1}a, left inset), phase matching is realized under the condition $\varphi_i=\omega_i D/v\sim \pi$.

The phase-matching concept is illustrated in Fig.~\ref{Fig1}, where we explore the coupling probability between an electron passing parallel and just outside a silver nanorod (Fig.~\ref{Fig1}a right inset) and the longitudinal plasmons supported by the particle. For fixed aspect ratio and electron energy, the coupling is maximized at an optimum value of the rod length $D$ (Fig.~\ref{Fig1}a), while the mode frequency redshifts with increasing $D$ due to retardation. To explore the validity of Eqs.~(\ref{quasieelsdep}) and (\ref{Fivarphii}), we consider the mode field distribution (Fig.~\ref{Fig1}b) and, in particular, its variation along the electron trajectory (Fig.~\ref{Fig1}c). Performing the Fourier transform in Eq.~(\ref{Fi}), we find a profile of $F_i(\varphi_i)$ (Fig.~\ref{Fig1}d) that matches very well the frequency-integrated excitation probability (Fig.~\ref{Fig1}e), whose maximum is in good agreement with the expected value $\varphi_i\sim\pi$.

\begin{figure*}
\centering
\includegraphics[width=1\textwidth]{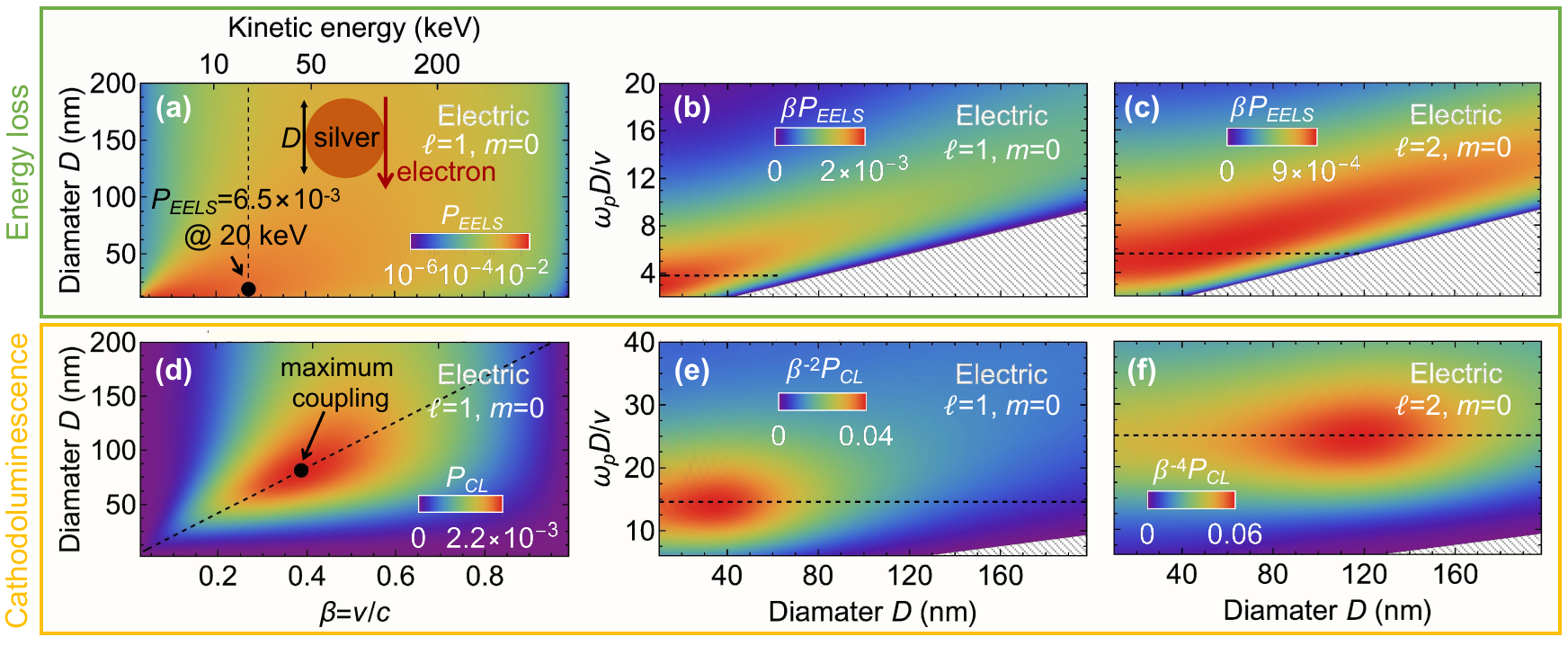}
\caption{{\bf Scaling of EELS and CL probabilities for silver spheres.} {\bf (a,b)}~Analysis of the frequency-integrated EELS probability (a) and the integrated probability multiplied by $\beta=v/c$ (b) for one of the electric dipole modes ($\ell=1$, $m=0$) excited by electrons passing grazing to a silver sphere (see inset in (a)). {\bf (c)}~Same as (b), but for an electric quadrupole mode ($\ell=2$, $m=0$). {\bf (d-f)}~Analogous analysis as in (a-c), but for CL. In panels (e) and (f), the probability is multiplied by $\beta^{-2}$ and $\beta^{-4}$, respectively, with $\beta=v/c$. The EELS probability is indicated by a label in (a) for $D=20$~nm and a KE of 20~keV. The maximum CL probability in (d) lies close to the dashed line $\wp D/v=9.8$ with $\hbar\wp=9.17$~eV (the Drude bulk frequency of silver \cite{JC1972}). Gray areas correspond to $v\geqslant c$.}
\label{Fig2}
\end{figure*}

Equation~\ref{quasieelsdep} corroborates the prediction made based on the quasistatic scaling of Eq.~(\ref{Peelsscaling}): the EELS coupling probability can be indefinitely increased by lowering the velocity while maintaining the phase-matching condition (i.e., the value of $\varphi_i$), such that $F_i$ takes its maximum value. In Fig.~\ref{Fig2}, we test this observation for a self-standing silver sphere (see supplementary Fig.~\ref{FigS1} for a gold sphere), calculated from the analytical solution in terms of Mie scattering for aloof electron excitation \cite{paper021}. Figure~\ref{Fig2}a shows that the coupling of a grazing electron to a dipolar plasmon can reach $\sim1\%$ for a KE of $20$ keV. Once we are in the quasistatic regime (size-independent mode frequency $\omega_i$), which is reached for a particle diameter $D\lesssim50$~nm (see Fig.~\ref{Fig2}b,c), we can increase the probability by reducing $D$, although in practice, a limit is imposed by nonlocal effects that shift and broaden the plasmon when $D$ is a few nm \cite{KV95,paper119,SKD12,RBW15} (i.e., not too large compared with the Fermi wavelength in the metal $\sim0.5$~nm).

As already mentioned in the derivation of Eq.~(\ref{Pclscaling}), the outcoupling process leading to CL emission introduces a dependence on the effective mode size and a quadratic scaling with the electron velocity. In addition, because the CL probability (Eq.~(\ref{cl})) is the square of an amplitude expressed as the sum of contributions from multiple modes, pairwise mode interference can emerge. For simplicity, we focus on modes with a small spectral overlap, such that any interference disappears, and then, we can rewrite the contribution of mode $i$ to the CL emission probability as
\begin{widetext}
\begin{align}
P_{{\rm CL},i}(\Rb_0,v)\approx \beta^2 \alpha  B_i \chi_i (\beta,\varphi_i)  \;\varphi_i^3F_i(\varphi_i)
{\rm [3D~quasistatic]},\label{quasicldep}
\end{align}
\end{widetext}
where $F_i(\varphi_i)$ and $\varphi_i$ are given by Eqs.~(\ref{Fivarphii}), while $\chi_i$ carries information on the radiative part of the process and the coefficient $B_i$ only depends on the material permittivity and its derivative at the mode frequency $\omega_i$ (see Appendix~\ref{AppendixE} for a detailed derivation). Although Eq.~(\ref{quasicldep}) has a similar structure as the coupling probability in Eq.~(\ref{quasieelsdep}), the presence of the emission function $\chi_i$ renders the scaling analysis of $P_{{\rm CL},i}$ more involved. However, $\chi_i$ becomes independent of $\beta$ and $\varphi_i$ for small particles (see Eq.~(\ref{chii}) in Appendix~\ref{AppendixE}), so that the CL probability vanishes as $\propto\beta^2$ with decreasing electron velocity. Therefore, in contrast to the EELS probability, which is favored by small velocities under phase-matching conditions, the maximum of CL emission must lie at some finite value of $\beta$.

Considering again silver nanospheres, $P_{{\rm CL},i}$ for $i$ corresponding to the electric ($\ell=1$) $m=0$ electric dipole exhibits an absolute maximum value of $\sim 0.2\%$ at $\beta\approx0.39$ ($\approx44$~keV) and $D\approx82$~nm (Fig.~\ref{Fig2}d). In addition, Fig.~\ref{Fig2}e,f corroborates the different behavior of the CL emission with electron velocity for different modes: $P_{{\rm CL},i}$ scales as $\beta^2$ for the dipole and $\beta^4$ for the quadrupole. Mode symmetry is therefore a critical aspect in the optimization of CL, and as a general rule, faster electrons interact more efficiently with more delocalized excitations.

\begin{figure*}
\centering
\includegraphics[width=0.67\textwidth]{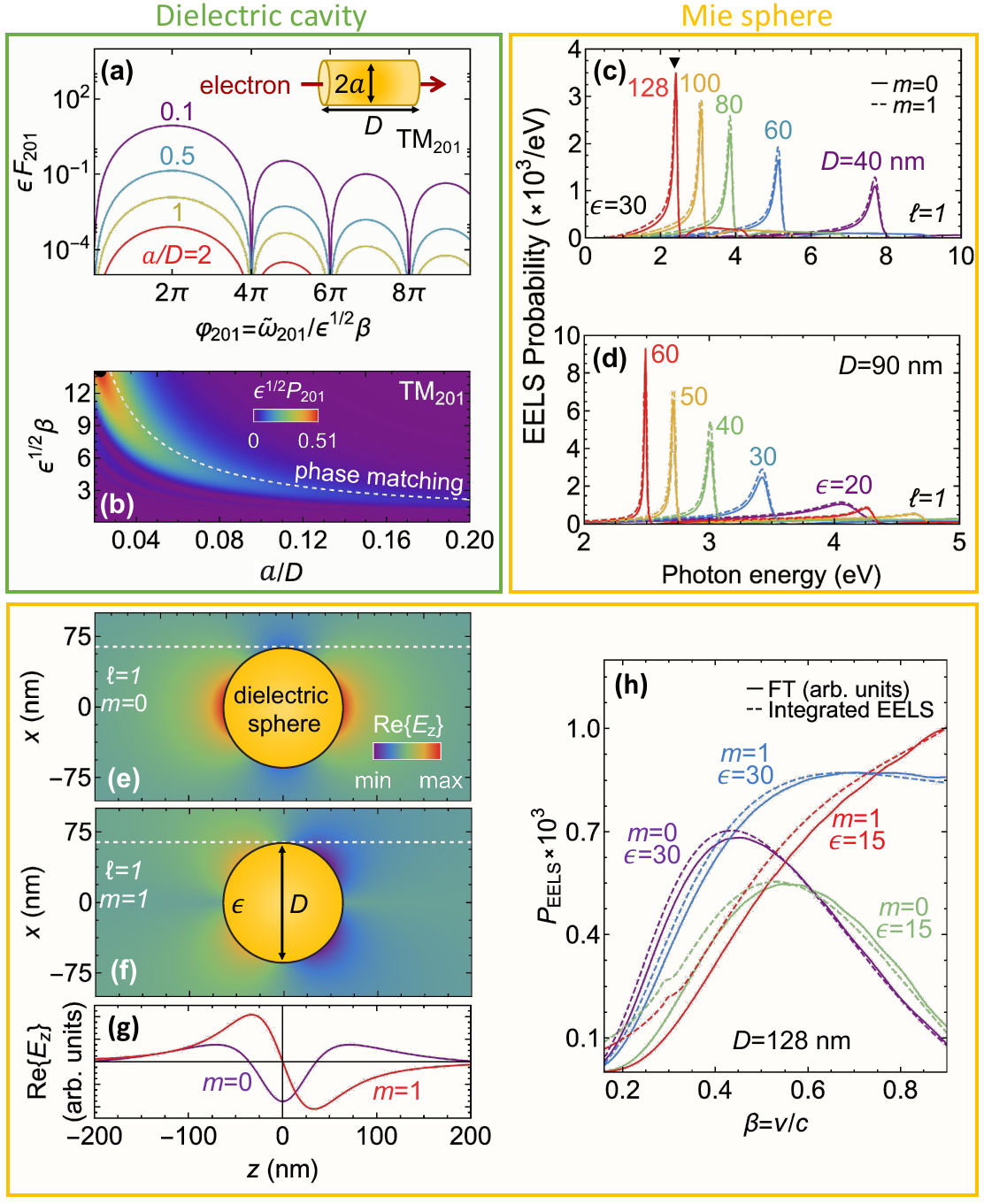}
\caption{{\bf Electron coupling to dielectric optical cavities.} {\bf (a,b)}~Coupling to a model cylindrical cavity (radius $a$, length $D$) coated by a perfect metal and filled with a material of permittivity $\epsilon$ (see inset in (a)). The electron is traversing the cavity with velocity $v$ along the axis. Panel (a) shows the squared spatial Fourier transform of the TM$_{201}$ mode field ($F_{201}$) along the axis as a function of the electron-velocity- and shape-dependent phase defined by $\varphi_{201}=\tilde{\omega}_{201}/\epsilon^{1/2}\beta$ with $\tilde{\omega}_{201}=\sqrt{(z_1D/a)^2+4\pi^2}$, where $z_1$ is the first zero of the Bessel function $J_0$. The mode excitation probability $P_{201}$ is proportional to $F_{201}$ (Eq.~(\ref{Peelscldiel})). Panel (b) shows the normalized probability $\epsilon^{1/2}P_{201}$ as a function of aspect ratio $a/D$ and $\sqrt{\epsilon}\beta=\sqrt{\epsilon}v/c$ (the electron velocity normalized to the speed of light in thein the dielectric). The phase-matching condition $\varphi_{201}=2\pi$ signals the probability maximum (dashed curve). {\bf (c-h)}~Excitation of a dielectric Mie sphere by a grazing electron. Panels (c,d) show the contribution of two electric dipolar modes ($\ell=1$ with azimuthal symmetries $m=0$ and 1) to the EELS spectra for different values of the particle diameter $D$ and permittivity $\epsilon$. Panels (e,f) present maps of the mode electric field amplitude (component along the e-beam direction $z$) for $D=180$~nm and $\epsilon=30$. Panel (g) represents the field as a function of position along the trajectory (dashed lines in (e,f)). Panel (h) compares the mode excitation probability (i.e., the frequency-integrated EELS) to the spatial Fourier transform of the field for $\epsilon=15$ (red and green curves) and $\epsilon=30$ (purple and blue curves, same conditions as in (g)).}
\label{Fig3}
\end{figure*}

\subsection{Dielectric Cavities}

For specimens of type (\textit{ii}) (lossless dispersionless permittivity $\epsilon$), neglecting radiative losses, one can also define a normalized field \cite{GL91} $\Eb_i(\rb)=\tilde{\Eb}_i(\rb/D)/D^{3/2}$ for each mode $i$ in terms of material- and size-independent functions $\tilde{\Eb}_i(\ub)$ of the scaled coordinate $\ub=\rb/D$, while the mode frequency scales as $\omega_i\propto1/D$ with size (see Appendices~\ref{AppendixC} and \ref{AppendixI} for details). Calculating the spatial Fourier transform of the field and the mode phase as prescribed by Eqs.~(\ref{Fivarphii}), the excitation probability is found to satisfy the scaling
\begin{align}
P_{{\rm EELS/CL},i}(\Rb_0,v)&\approx2\pi\frac{\alpha}{\beta}  F_i(\varphi_i) \quad {\rm [3D~dielectrics]}.\label{Peelscldiel}
\end{align}
A rigorous description including radiative losses requires a proper treatment of the excited modes as resonances (e.g., through quasinormal modes \cite{CLM98,GH15,LYV18,FHD19}), which should produce just small corrections for high-quality resonances as those considered below. Equation~(\ref{Peelscldiel}) can also be derived by inserting the mode decomposition of the Green tensor (Eq.~(\ref{GreenEiEi}) in Appendix~\ref{AppendixC}) into Eq.~(\ref{eels}), separating the contribution of a single mode $i$, and integrating over frequency.

We start our discussion by considering an e-beam traversing a cylindrical dielectric cavity (permittivity $\epsilon$, radius $a$, length $D$, see inset in Fig.~\ref{Fig3}a) along the axis. Although this model system has mainly academic interest, it captures the general characteristics of dielectric cavities as well as their interaction with fast electrons, so we hope that it can help in general when investigating any type of cavities in which light is trapped in the form of standing waves bouncing at the boundaries (in contrast to, for example, small plasmonic cavities, where the response of the material provides the leading trapping mechanism). In this system, Eq.~(\ref{Peelscldiel}) is exact. For simplicity, we assume the dielectric to be coated by a perfect electric conductor. Then, the dependence on geometry enters only through the aspect ratio $D/a$. The cavity supports transverse electric and magnetic (TE and TM) modes, but only the latter have a nonzero electric field component along the e-beam and can therefore be excited. Modes are further indexed by longitudinal, azimuthal, and radial numbers $n$, $m$, and $k$, respectively. For the axial trajectory under consideration, only $m=0$ modes couple to the electron. We focus in particular on the $n=2$ and $k=1$ TM mode (i.e., TM$_{201}$). The corresponding function $F_i=F_{201}$ entering Eq.~(\ref{Peelscldiel}) admits a closed-form analytical expression (Eq.~(\ref{FCCnmk}) in Appendix~\ref{AppendixJ}), while the mode frequency reduces to $\omega_{201}=(c/D\sqrt{\epsilon})\sqrt{(z_1D/a)^2+4\pi^2}$, where $z_1\approx2.405$ is the first zero of the $J_0$ Bessel function.

We find that $F_{nmk}$ increases with the square of the aspect ratio $D/a$ and presents an absolute maximum at $\varphi_{nmk}\approx n\pi$ as a function of the coupling phase, signaling the phase-matching condition for this system (see Fig.~\ref{Fig3}a). Moving along the phase-matching curve requires changing the permittivity while modifying the cylinder aspect ratio (dashed curve in Fig.~\ref{Fig3}b). Upon inspection of Eq.~(\ref{FCCnmk}) in Appendix~\ref{AppendixD}, we find that the excitation probability scales as $P_{nmk}\propto D/(a\sqrt{\epsilon})$ for $D/a\gg n\pi$, thus producing stronger coupling for large aspect ratios, in agreement with Fig.~\ref{Fig3}b, where we observe moderate probabilities ($P_{201}\sim3\%$) even for ultrarelativistic electrons ($\sim300$~keV), high permittivities ($\epsilon\sim100$), and extreme aspect ratios ($D/a\sim25$).

A paradigmatic configuration consists of an electron passing grazingly to a self-standing dielectric sphere (diameter $D$, permittivity $\epsilon$) supporting Mie resonances (see Appendix~\ref{AppendixG}). The EELS probability can be calculated analytically using closed-form expressions derived from Mie theory \cite{paper021,paper149}. For illustration, we consider electric dipolar modes ($\ell=1$) with either $m=0$ or 1 azimuthal symmetry, whose associated electric fields are plotted in Fig.~\ref{Fig3}e,f. The EELS probability is then given by $\Gamma_{{\rm EELS}}(\Rb_0,v,\omega)=(3\alpha g_m/\pi\omega\beta^4\gamma^4)K_m^2(\omega R_0/v\gamma){\rm Im}\{t_1^E(\omega)\}$ with $R_0=D/2$ (grazing incidence), where $K_m$ are modified Bessel functions, $t_1^E(\omega)$ is the electric dipole scattering coefficient \cite{paper021}, $g_0=2$, and $g_1=\gamma^2$. (See supplementary Fig.~\ref{FigS3} for a comprehensive study of excitation probabilities of modes with different symmetries.) Like in the cylinder, the resonant frequency decreases with increasing permittivity and cavity size (see EELS spectra in Fig.~\ref{Fig3}c,d). This behavior is consistent with the high-permittivity frequency scaling $\omega_i\propto c/2\sqrt{\epsilon} D$ (see supplementary Fig.~\ref{FigS2}a), suggesting that the optimization of the coupling probability based on phase-matching arguments could benefit from using materials with high permittivities combined with slow electrons. However, this approach is incompatible with the observed reduction in the spectrally integrated loss function $\mathcal{I}^E_1=\int_i d\omega\,{\rm Im}\{ t_1^E(\omega)\}/\omega_i\propto1/\epsilon^2$ as $\epsilon$ increases (see supplementary Fig.~\ref{FigS2}b). Instead, we should consider phase-matching with cavities of moderate permittivity and using large electron velocities (see supplementary Fig.~\ref{FigS3}, where relativistic corrections are observed to contribute favorably). For example, we find an optimum coupling of $\sim 0.1$\% for $\epsilon=15$ and $\beta>0.8$ in the spheres of Figs.~\ref{Fig3}h and S3.

In summary, the scaling $P_{{\rm EELS},i}\propto F_i(\varphi_i)/\beta$ for dielectric structures suggests that optimization of the coupling to electrons can be pursued by playing with the shape of the cavity while adjusting the size and the electron velocity such that the phase $\varphi_i$ is kept constant at a value that satisfies the geometry-dependent phase-matching condition. In addition, we note that increasing $\epsilon$ to control the effective mode size leads in general to a reduction in the coupling, unless the electron--cavity interaction time can be increased by changing the morphology of the structure (e.g., in an elongated cylinder), although this strategy is limited in practice by the maximum currently available permittivities $\epsilon\sim20$ at near-infrared/visible frequencies \cite{TSR22}.

\subsection{Coupling to 2D Structures}

Two-dimensional (2D) materials have become a relevant ingredient in nanophotonics because of their robustness, flexible integration, large tunability through electrical gating, and extraordinary optical properties that include a plethora of long-lived polaritons in van der Waals materials \cite{paper283} such as plasmons in graphene \cite{GPN12,paper235,NMS18}, phonon-polaritons in hexagonal boron nitride \cite{GDV18} (hBN) and $\alpha$-MoO$_3$ \cite{MAL18,paper396}, and excitons in transition-metal dichalcogenides \cite{LCZ14} (TMDs), as well as plasmons in atomically thin noble metal nanostructures \cite{paper335,paper427}. The scaling properties of 2D nanostructures are different from those of 3D particles, as one is interested in maintaining a constant thickness $d$ while varying the lateral size $D$. In addition, the mode frequencies of atomically thin structures are generally small, so they can be described in the quasistatic limit.

\begin{table*}
\centering
\includegraphics[width=0.7\textwidth]{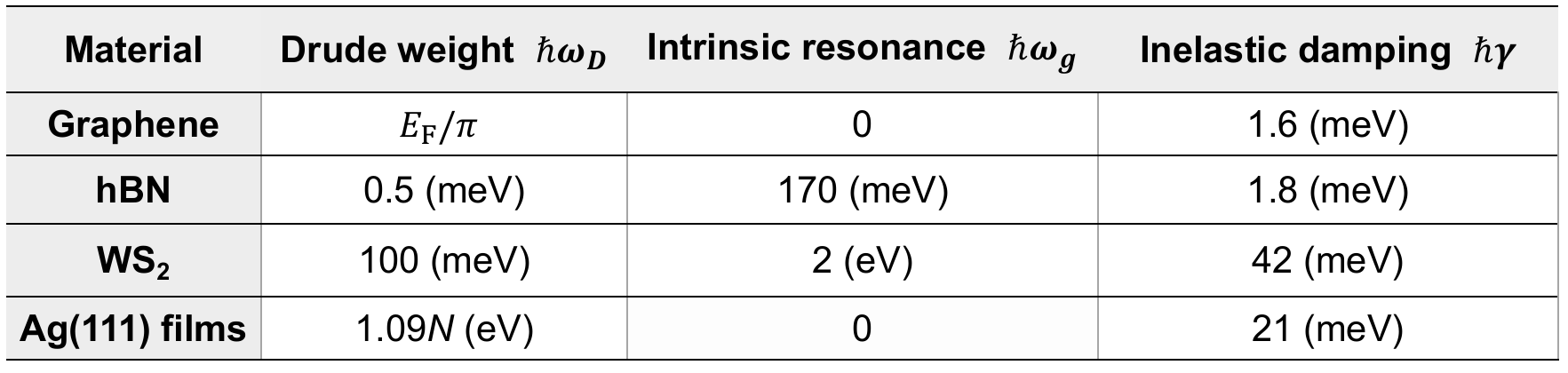}
\caption{Parameters used to compute the 2D conductivity $\sigma(\omega)=(\ii e^2/\hbar )\,\omega_D/(\omega-\omega_g-\ii \gamma)$ (Eq.~(\ref{sigmaD})) in the calculations of the coupling probabilities of Fig.~\ref{Fig4} as well as for Ag(111) in Fig.\ \ref{FigS4}. We focus on plasmons in graphene doped to a Fermi energy $\EF$, the LO phonon of hBN (data from ref~\citenum{paper361}), the A exciton of monolayer WS$_2$ (ref~\citenum{LCZ14}), and plasmons in Ag(111) films consisting of $N$ atomic planes, where\cite{paper335} $\omega_D=\hbar\wp^2Nd_z/4\pi e^2$ involves the bulk plasmon energy\cite{JC1972} $\hbar\wp=9.17$~eV and the atomic-layer spacing $d_z\approx0.236$~nm. Gold films share these parameters, but the damping changes to \cite{JC1972} 71~meV.}
\label{Table1}
\end{table*}

The zero-thickness approximation (ZTA) is commonly employed to reliably describe 2D materials assuming $d\to0$ and using a 2D conductivity $\sigma(\omega)$. This approach renders similar results as those obtained for films of finite thickness $d\ll D$ described through a permittivity $\epsilon(\omega)\approx4\pi\ii\sigma(\omega)/\omega d$ \cite{paper235}. Polaritons in 2D materials are generally well-described in terms of local, frequency-dependent conductivities of the form
\begin{align}
\sigma(\omega)=\frac{\ii e^2}{\hbar}\,\frac{\omega_D}{\omega-\omega_g+\ii \gamma}, \label{sigmaD}
\end{align}
with three parameters defined as a Drude weight $\omega_D$, an intrinsic-resonance frequency $\omega_g$, and a phenomenological inelastic damping rate $\gamma$. Typical values for these parameters are provided in Table~\ref{Table1} for materials of interest (graphene, hBN, a TMD, and $N$ layers of Ag(111)). Adopting a 2D quasistatic mode expansion, we find morphology-dependent mode frequencies $\omega_i$ prescribed by characteristic scale-invariant eigenvalues $\eta_i$ through the condition \cite{paper228,paper235} $\ii\sigma(\omega_i)/\omega_i=D\eta_i$ (see Appendix~\ref{AppendixB}). Plugging Eq.~(\ref{sigmaD}) into this expression, we find
\begin{align}
\omega_i\approx\omega_g/2+\sqrt{\omega_g^2/4-e^2\omega_D/\hbar\eta_i D}-\ii\gamma/2, \label{wi2D} 
\end{align}
which contains a common imaginary part (i.e., equal mode decay rate independent of size and morphology). Under the assumption of long-lived polaritons, we neglect mode damping in what follows and only retain the real part of $\omega_i$.

When the structure size is scaled by a factor $\mu$, using Eq.~(\ref{eelsquasi}) in combination with Eq.~(\ref{2dWscaling}), we find that the scaling properties of the EELS and CL probabilities follow the same relations as in Eqs.~(\ref{Peelsscaling1}) and (\ref{Pclscaling}), respectively, but modified to accommodate a change in the surface conductivity according to
\begin{widetext}
\begin{align}
&\tilde{P}_{{\rm EELS},i}(\Rb_0,v,\sigma)=\frac{1}{\mu}\,P_{{\rm EELS},i}(\Rb_0/\mu,v/\mu,\sigma/\mu) &{\rm [2D~quasistatic]}, \nonumber\\
&\tilde{P}_{{\rm CL},i}(\Rb_0,v,\sigma)\approx \mu^2\,P^{(0)}_{{\rm CL},i}(\Rb_0/\mu,v/\mu,\sigma/\mu)+\mu^3\,P_{{\rm CL},i}^{(1)}(\Rb_0/\mu,v/\mu,\sigma/\mu)+\cdots & {\rm [2D~quasistatic]} \nonumber
\end{align}
(see Appendix~\ref{AppendixF} for details).

Using the noted quasistatic mode expansion, the probability in Eq.~(\ref{eels}) leads to (see Appendix~\ref{AppendixF})
\begin{align}
P_{{\rm EELS},i}(\Rb_0,v)&\approx \frac{\alpha^2}{\beta^2}\frac{\omega_D}{2|\eta_i|\omega_i} \,\frac{1}{\varphi_i}F_i\left(\varphi_i\right) \quad\quad {\rm [2D~quasistatic]}, \label{2dpeels}
\end{align} 
\end{widetext}
where $F_i(\varphi_i)$ and $\varphi_i$ are again given by Eqs.~(\ref{Fivarphii}). Comparison between Eqs.~(\ref{quasieelsdep}) and (\ref{2dpeels}) reveals that the dimensionality of the system radically affects its coupling to free electrons: in 2D nanostructures, $P_{{\rm EELS},i}\propto 1/\beta^2$; in contrast, the scaling goes like $1/\beta$ in 3D systems. Another important consequence of Eq.~(\ref{2dpeels}) relates to the behavior of $\omega_i$ when modifying the particle size $D$ (see Eq.~(\ref{wi2D})): in materials with $\omega_g=0$ such as graphene, $\omega_i\propto 1/\sqrt{D}$ guarantees a coupling maximization through a combined reduction of size $D$ and electron velocity $\beta$ (see below). In the presence of an intrinsic material resonance ($\omega_g\neq0$), the behavior of $P_{\rm EELS}$ closely resembles the one of a 3D system because any variation of $\omega_i$ with size can be neglected as long as $D|\eta_i|\gg e^2\omega_D/\hbar \omega_g^2$ (e.g., for $D|\eta_i|\gg 0.04$ nm in WS$_2$, see Table~\ref{Table1}). This condition is always met in the configurations explored in this work.


\begin{figure*}
\centering
\includegraphics[width=0.85\textwidth]{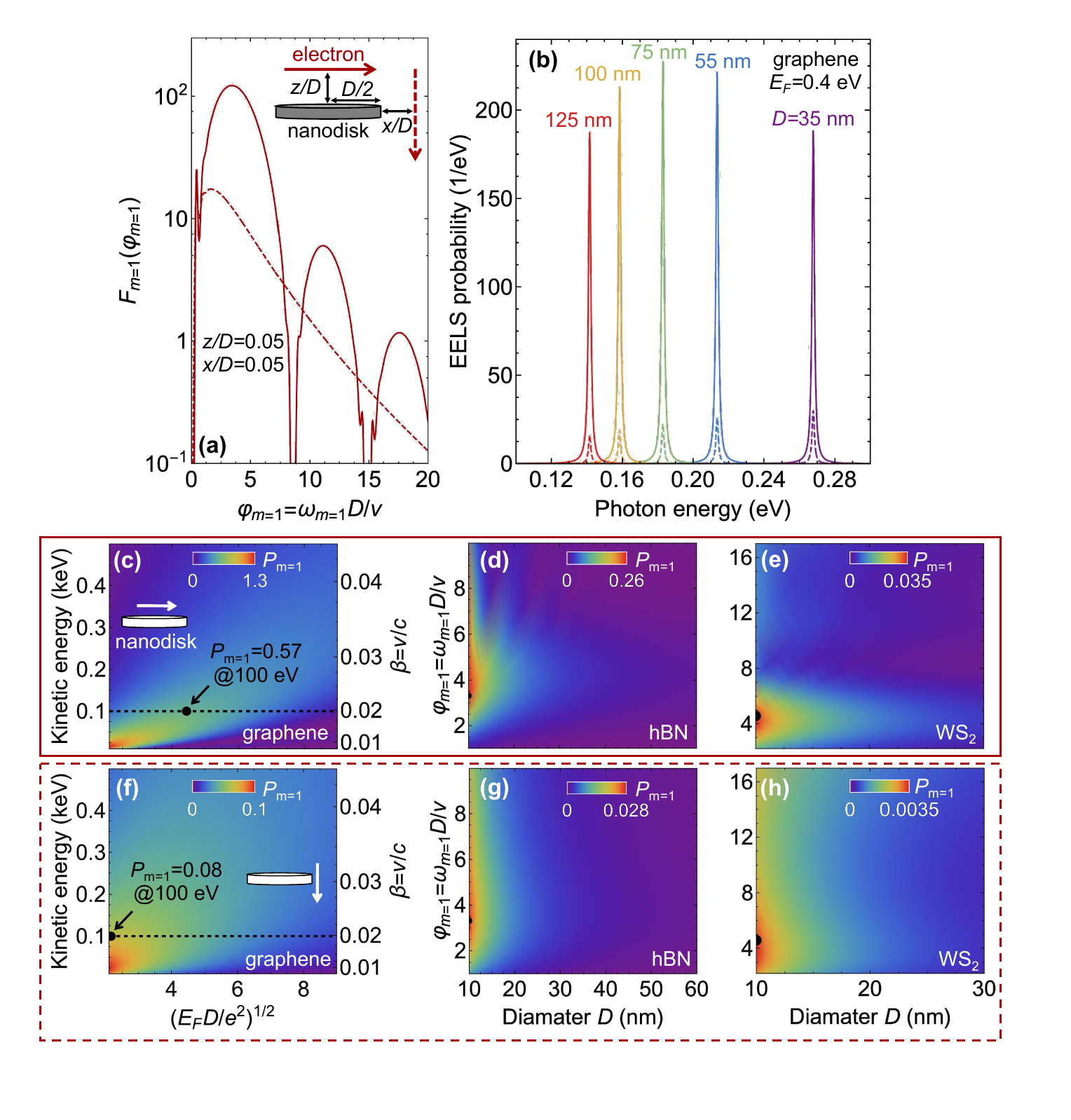}
\caption{{\bf Electron coupling to 2D nanodisks.} {\bf (a)}~Dimensionless scaling function $F_{m=1}$ (see Eq.~(\ref{2dpeels})) for an electron passing parallel (solid curve) or perpendicular (dashed curve) to a nanodisk of diameter $D$ supporting a mode with azimuthal number $m=1$ and frequency $\omega_{m=1}$. The normalized impact parameter is $z/D=0.05$ for the parallel trajectory and $x/D=0.05$ for the perpendicular one (see inset). {\bf (b)}~Energy-loss spectra for a 100~eV electron moving parallel (solid curves) or perpendicular (dashed curves) to a graphene nanodisk doped to a Fermi energy $\EF=0.4$~eV for different disk sizes (see color-matched labels). {\bf (c,f)}~Frequency-integrated probability in graphene nanodisks as a function of electron KE and the dimensionless parameter $(\EF D/e^2)^{1/2}$. Dashes lines indicate a KE of 100~eV and black dots signal the maximum coupling probability at such energy. {\bf (d,g)}~Same as (e,h), but for hBN nanodisks as a function of $D$ and $\varphi_{m=1}\equiv\omega_{m=1}D/v$. {\bf (e,h)}~Same as (d,g), but for WS$_2$ nanodisks. Panels (c-e) and (f-h) correspond to the parallel and perpendicular electron trajectories, respectively (see inset in (a)).}
\label{Fig4}
\end{figure*}

We illustrate the predictions of Eq.~(\ref{2dpeels}) by considering a 2D self-standing nanodisk of diameter $D$ excited by an electron passing either parallel or perpendicular to the surface (see inset in Fig.~\ref{Fig4}a). We focus on the lowest-order dipole mode with $m=1$ azimuthal symmetry, corresponding to an eigenvalue $\eta_{m=1}=-0.0731$ and an associated charge density $\sigma_{m=1}(\Rb)\underset{\sim}{\propto}(R/D)\big[1+\ee^{-5(1-2R/D)}/\left(4\sqrt{1-2R/D}\right)\big]\cos\varphi_\Rb$ for $\Rb=(x,y)$ in the disk (see ref~\citenum{paper254}). From $\sigma_{m=1}(\Rb)$, we obtain the mode field $\tilde{\Eb}_{m=1}(\ub)$ (see Appendix~\ref{AppendixB}), and in turn, using Eq.~(\ref{Fi}), the $F_{m=1}(\varphi_{m=1})$ coefficient that enters Eq.~(\ref{2dpeels}). We find that $F_{m=1}(\varphi_{m=1})$ (see Fig.~\ref{Fig4}a) displays an absolute maximum around $\varphi_{m=1}\approx4$ for the parallel trajectory (solid red curve) as well as an oscillatory behavior produced by spatial modulation of the charge density around the disk, whereas $F_{m=1}(\varphi_{m=1})$ decreases monotonically with phase for the perpendicular e-beam (dashed red curve). Applying these functions to compute the corresponding EELS spectra, we obtain the results presented in Fig.~\ref{Fig4}b for graphene nanodisks of different sizes doped to a Fermi energy $\EF=0.4$~eV. We then integrate the spectra over the observed peaks to calculate the mode excitation probabilities presented in  Fig.~\ref{Fig4}c,f as a function of electron kinetic energy and scaled size. Upon inspection of the analytical expressions for the excitation probability $P_{m=1}$, we corroborate that it depends on Fermi energy and size only through the dimensionless parameter $\EF D/e^2$ in this material \cite{paper228}.  We also perform a similar analysis for hBN (Fig.~\ref{Fig4}d,g) and WS$_2$ (Fig.~\ref{Fig4}c,f) disks using the conductivity parameters listed in Table~\ref{Table1} (see also results for ultrathin Ag(111) nanodisks in supplementary Fig.~\ref{FigS4}). Among these materials, graphene stands out for its ability to provide very high electron--sample coupling probabilities ($\sim60\%$ for 100~eV parallel electrons, a diameter $D\sim 73$~nm, and a Fermi energy $\EF=0.4$~eV). This needs to be compared to the maxima of $\sim26\%$ and $\sim4\%$ observed in hBN and WS$_2$ disks, respectively, for $D=10$~nm. We attribute this advantage of graphene to its scaling $\varphi_{m=1}\propto\sqrt{D}$, in contrast to materials hosting intrinsic resonances, which exhibit a $\varphi_{m=1}\propto D$ behavior. In consequence, small graphene disks excited by electrons with low velocities undergo high excitation probabilities that cannot be reached with the other two materials.

The frequency-integrated CL probability associated with 2D systems is obtained by following a similar procedure as used to derive Eq.~(\ref{quasicldep}). We find $P_{{\rm CL},i}(\Rb_0,v)\approx (\alpha^{5}/\beta^2)C_i  \chi_i F_i(\varphi_i)/\varphi_i$ (see Eq.~(\ref{tmp6}) in Appendix~\ref{AppendixF}), where the coefficient  $C_i=\omega_D^4/8\pi \eta_i^2 \gamma\omega_i[(\omega_i-\omega_g)^2+\gamma^2]$ is weakly dependent on $D$, while the phase $\varphi_i$ is still given by Eq.~(\ref{varphii}). Interestingly, since the emission process involves the particle surface rather than the volume, the scaling of the CL emission intensity significantly departs from the one observed in 3D particles ($\propto\beta^2$ in the small-particle limit, see Eq.~(\ref{quasicldep})).

\begin{figure*}
\centering
\includegraphics[width=1.0\textwidth]{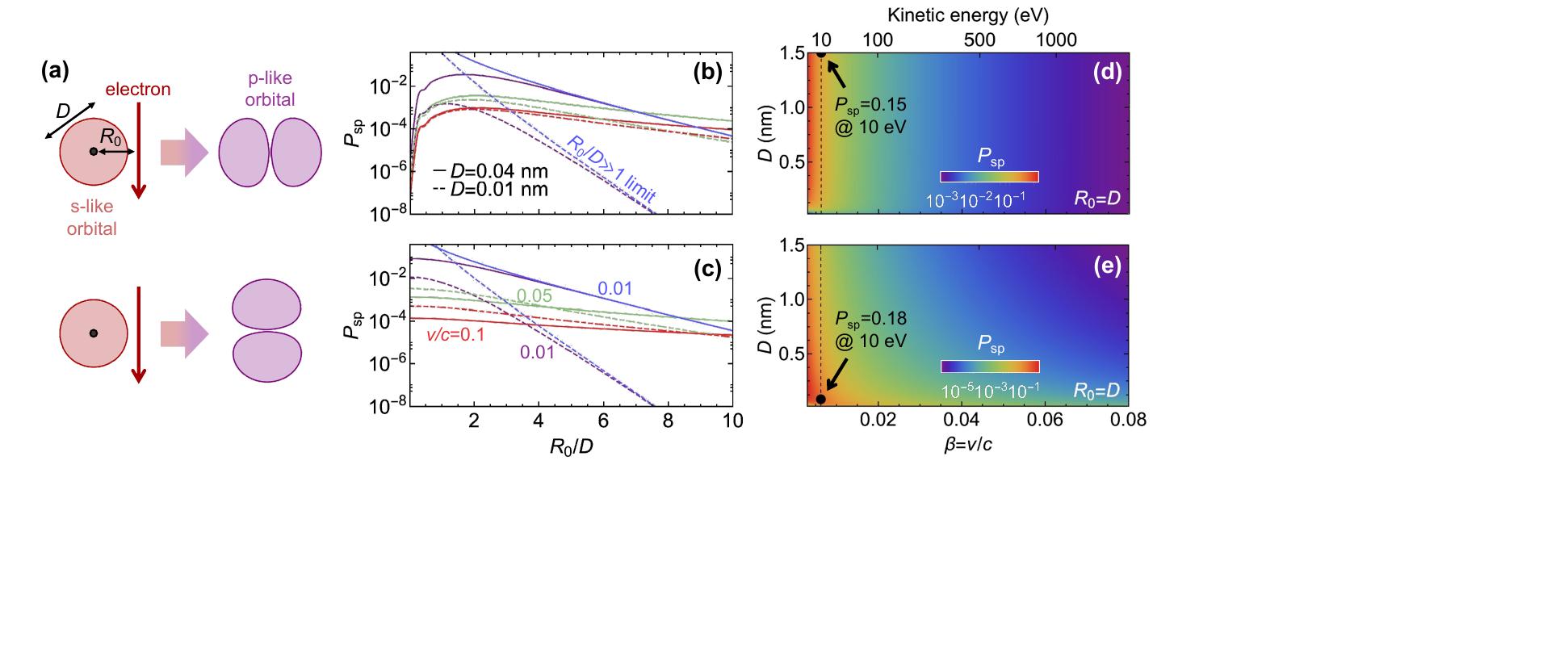}
\caption{{\bf Electron-induced atomic-like s-p transition.} {\bf (a)}~Illustration of a bound-electron transition from an s orbital of size $D$ to a p orbital triggered by the interaction with a free electron passing at a distance $R_0$ from the origin of such orbitals. We consider final p orbitals oriented either perpendicularly or parallel to the electron trajectory (upper and lower sketches, respectively). {\bf (b,c)}~Excitation probability (from s to p) as a function of the scaled impact parameter $R_0/D$ for p orbitals with perpendicular (b) or parallel (c) orientations. Dashed and solid curves are obtained for two different orbital sizes, curve colors refer to different electron velocities $\beta=v/c$ (see labels), and the $R_0\gg D$ limit is also shown for $v/c=0.01$. {\bf (d,e)}~Excitation probability under the same conditions as in (b,c) as a function of electron velocity/energy (lower/upper horizontal scales) and orbital size for $R_0=D$. Dashed vertical lines indicate an electron KE of 10~eV, while black dots signal the maximum coupling probability at such energy.}
\label{Fig5}
\end{figure*}

\subsection{Free-Electron Coupling to Atomic-Like Excitations}

The observation of single-atom electronic transitions in the optical regime has remained a challenge in electron microscopy. In a relevant study, the modifications produced in the plasmonic response by the presence of atomic defects have been monitored through EELS in nanomaterials \cite{ZLN12}, while CL emission from defects in nanodiamonds has also been resolved \cite{MTC15}. The atomic-like excitations supported by quantum dots \cite{BCK99}, excitons in semiconductors \cite{K1963,LCZ14}, and defects in TMDs \cite{paper354} are promising candidates for probing the interaction at the single-electron level, so we are interested in finding the optimum excitation probability in these types of systems. To this end, we explore a simple model consisting of a single electron bound to an effective Coulomb potential in an initial s orbital $\psi_s(\rb)=\ee^{- r/D}/\sqrt{D^3\pi}$, and study the transition to a p orbital $\psi_p(\rb)=(\rb\cdot\nn)\ee^{-r/2D}/\sqrt{(2D)^5\pi}$ (oriented along the direction of a unit vector $\nn$ either parallel or perpendicular to $\vb$, see Fig.~\ref{Fig5}a) induced by the passage of a free electron moving with velocity $\vb=v\zz$ and impact parameter $\Rb_0$ relative to the Coulomb singularity. The size of the atomic-like states is defined by $D$, which also determines the transition energy $\hbar\omega_{sp}=3\hbar^2/(8\me D^2)$ (see Appendix~\ref{AppendixK}).

In analogy to the previously studied systems, we express the excitation probability in terms of a retarded electric field produced by the bound-electron current when transitioning from the ground state to the excited state. The spatial profile of that electric field is captured by a dimensionless vector field $\tilde{\Eb}_{sp}(\rb/D)$ (see Appendix~\ref{AppendixK}) that allows us to write the excitation probability as
\begin{align}
P_{sp}(\Rb_0,v)= \frac{\alpha^2}{\beta^2}\frac{1}{\varphi_{sp}}F_{sp}(\varphi_{sp}), \label{Psp}
\end{align}
where $F_{sp}(\varphi_{sp})$ is given by Eq.~(\ref{Fi}) (see Eq.~(\ref{Fsptmp1}) for an explicit expression) and depends on the phase $\varphi_{sp}=\omega_{sp}D/v$ (see detailed derivation in Appendix~\ref{AppendixK}). As in the nanoparticles discussed above, we note again that $F_{sp}(\varphi_{sp})$ is the squared Fourier transform of the scaled transition field. In the quasistatic limit, the field $\tilde{\Eb}_{sp}(\rb/D)$ becomes independent of the size of the orbital $D$, which enters only via the phase $\varphi_{sp}\propto1/D$. We find that $P_{sp}$ is maximized for $R_0/D\sim 1.4$ when the electron moves perpendicular to $\nn$, whereas it saturates to a finite value as $R_0\rightarrow0$ when $\vb\parallel\nn$ (see Fig.~\ref{Fig5}b). Because $\omega_{sp}\propto1/D^2$, Eq.~(\ref{Psp}) prescribes that $P_{sp}$ only depends on $\beta=v/c$ and the orbital size $D$. We can thus present universal plots for the excitation probability (Fig.~\ref{Fig5}d,e), revealing values as high as $\sim15$\% for 10~eV electrons and $D\sim1.5$~nm in the perpendicular orientation (see Fig.~\ref{Fig5}d).

In the long-distance limit ($R_0\gg D$), the electron only sees a dipolar transition similar to those in metal spheres. This regime is already approached for $R_0/D\gtrsim 6$ (Fig.~\ref{Fig5}b,c). Because the transition dipole scales as $d\propto D$ and the excitation probability as $P_{sp}\propto d^2\omega_{sp}^2K_m^2(\omega_{sp}R_0/v\gamma)$ with $m=0$/1 for $\nn$ oriented parallel/perpendicular to $\vb$ (see a detailed expression in Appendix~\ref{AppendixK}), the perpendicular orientation produces a larger coupling (cf. Figs.~\ref{Fig5}d and \ref{Fig5}e). Furthermore, under the expected condition that the size of the system is small compared to the transition wavelength (i.e., $\omega_{sp}D/c\ll1$), $P_{sp}$ becomes approximately independent of $\omega_{sp}$ and scales with the size of the system as $D^2$ if $R_0\gg D$.

\begin{figure*}
\centering
\includegraphics[width=1.0\textwidth]{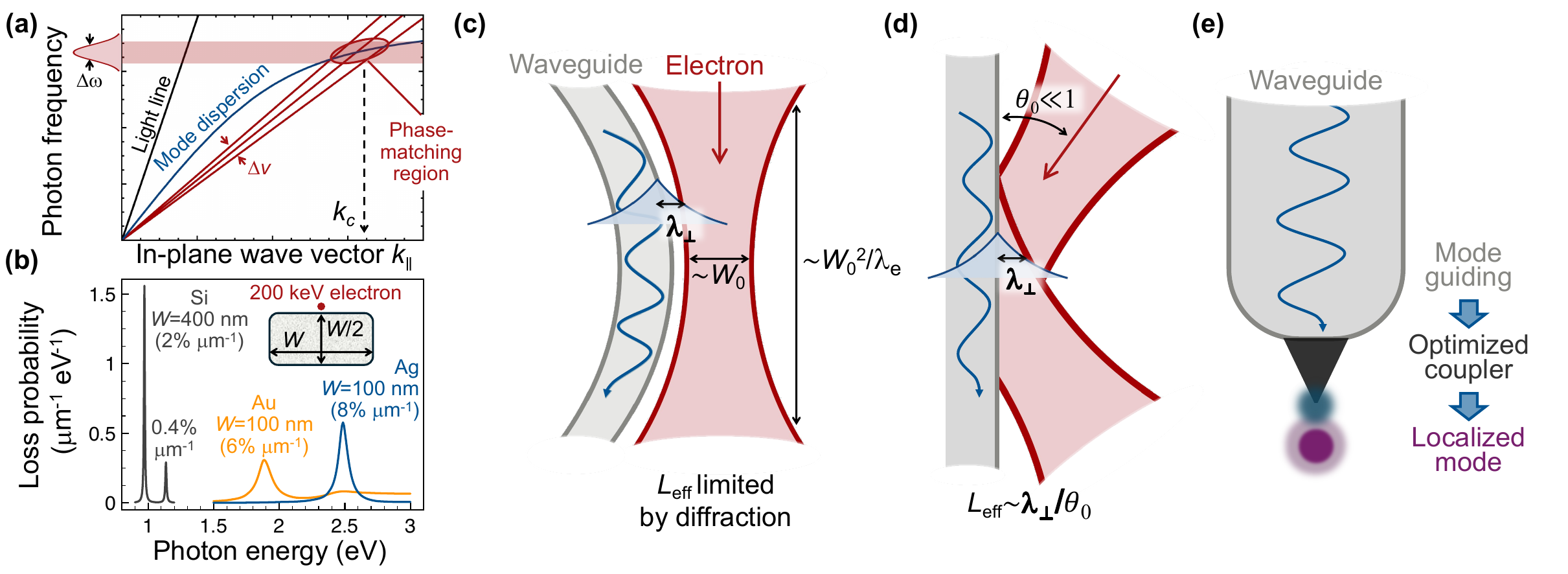}
\caption{{\bf Excitation of waveguide modes by free electrons.} {\bf (a)}~An electron moving parallel to the waveguide can efficiently excite guided modes when the phase-matching line $\omega=\kpar v$ (red), which relates the in-plane wave vector $\kpar$ and frequency $\omega$ that the electron can transfer, crosses the mode dispersion relation of the waveguide (dark blue) at a wave vector $\kpar=k_c$. The uncertainty in the parallel electron velocity $v$ (e.g., through the velocity spread $\Delta v$ of the e-beam) translates into a finite width $\Delta\omega$ of the generated optical spectrum. {\bf (b)}~The excitation probability reaches values of a few percent per micron of electron path length for metallic and dielectric waveguides with commonly used dimensions, here illustrated for gold, silver, and silicon waveguides with rectangular cross sections, as indicated in the inset. We show the corresponding loss spectra and the mode-integrated probabilities (in parentheses). Waveguide edges are rounded with a radius of curvature of 10~nm. The electron is passing 1~nm away from the surface at the position indicated by the dot in the inset. We take $\epsilon=12+0.1\,\ii$ for Si and $\epsilon(\omega)$ from ref~\citenum{JC1972} for the metals. {\bf (c)}~Large interaction lengths can be realized in curved waveguides matching the profile of a Gaussian e-beam. The penetration distance $\lambda_\perp$ of the waveguide mode amplitude outside the surface relative to the e-beam waist $W_0$ determines the effective interaction length $L_{\rm eff}$. {\bf (d)}~In a simpler configuration, a broad e-beam can be specularly reflected at the surface of a planar waveguide upon incidence with a glancing angle $\theta_0\ll1$, thus defining an effective interaction length $L_{\rm eff}\approx\lambda_\perp/\theta_0$. {\bf (e)}~The so-generated waveguide modes can be efficiently coupled to localized excitations using integrated optics schemes.}
\label{Fig6}
\end{figure*}

\subsection{Excitation of Waveguide Modes}

We conclude by discussing how waveguide modes confined in one or two spatial dimensions can be efficiently excited by electrons moving parallel to the propagation direction under phase-matching conditions \cite{paper180,K19,HRF21,FHA22,DDB23}. In the nonrecoil approximation, assuming a structure with translational invariance along the e-beam direction, electrons are capable of exciting modes that have a longitudinal wave vector $\kpar$ and frequency $\omega$ within the $\omega=\kpar v$ line determined by the electron velocity $v$ \cite{paper149}. Waveguide modes can thus be excited when this line crosses their dispersion relation $\omega=\omega_{\kpar}$ at a point with wave vector $\kpar=k_c$, as illustrated in Fig.~\ref{Fig6}a. We limit our analysis to one-dimensional (1D) waveguides, where the generated modes are easier to collect. This configuration was suggested as a basis to generate single photons heralded by electron energy losses \cite{paper180} and demonstrated in a recent experiment \cite{FHA22}.

We find it convenient to define an excitation probability per unit of electron path length $dP_{\rm EELS}/dz$, which is derived in Appendix~\ref{AppendixL} starting from Eq.~(\ref{Peelscldiel}) and can be expressed as
\begin{widetext}
\begin{align}
\frac{dP_{\rm EELS}(\Rb_0)}{dz}=\frac{2\pi\alpha c}{\omega_{k_c}(1-v_{k_c}/v)}\big|E_{k_c,z}(\Rb_0)\big|^2 \quad \quad {\rm [1D~waveguide]}, \label{Pwaveguide}
\end{align}
\end{widetext}
where $\omega_{k_c}$ and $v_{k_c}=\partial\omega_{k_c}/\partial k_c$ are the frequency and group velocity of the phase-matching mode with parallel wave vector $\kpar=k_c$, and $\Eb_{k_c}(\Rb_0)$ is the corresponding mode field evaluated at the transverse position of the e-beam $\Rb_0$. For waveguides made of a lossless, nondispersive material, the field satisfies the normalization condition $\int d^2\Rb\,\epsilon(\Rb)|\Eb_{k_c}(\Rb)|^2=1$, with $\epsilon(\Rb)=1$ outside the waveguide and equal to the material permittivity inside it; then, the factor $\big|E_{k_c,z}(\Rb_0)\big|^2$ in Eq.~(\ref{Pwaveguide}) scales as $1/\mu^2$ when the cross section of the waveguide and $\Rb_0$ are both enlarged by a factor $\mu$, and therefore, the loss rate in the scaled system ($d\tilde{P}_{\rm EELS}/dz$) is related to the original one ($dP_{\rm EELS}/dz$) through $d\tilde{P}_{\rm EELS}(\Rb_0)/dz=\mu^{-2}dP_{\rm EELS}(\Rb_0/\mu)/dz$. For waveguides with a dispersive permittivity, mode normalization involves contributions from magnetic and electric terms \cite{BS06_2}, generally leading to a more complex scaling behavior. Nevertheless, in plasmonic waveguides with highly confined modes, the electric component dominates and one can use the above normalization condition with $\epsilon(\Rb)$ replaced by $\partial[\omega\epsilon(\Rb,\omega)]/\partial\omega$ (evaluated at $\omega=\omega_{k_c}$), so that the scaling with $\mu$ remains the same as for dielectric waveguides.

Figure~\ref{Fig6}b shows examples of the excitation of plasmons in gold and silver waveguides as well as propagating optical modes in silicon waveguides using grazingly incident 200~keV electrons. We consider rectangular waveguides with dimensions that are feasible using currently available nanofabrication techniques. The excitation probability $dP_{\rm EELS}/dz$, normalized per unit of electron path length and integrated over the spectral peak of the mode, lies in the range of a few percent per micron. Over the entire electron trajectory, the total excitation probability is then given by this quantity multiplied by the effective electron interaction length $L_{\rm eff}$. In these calculations, we place the electron very close to the material surface (electron--surface distance $x=1$~nm), but analogous results are obtained at larger distances $x$, for which the probability is reduced by a factor $\ee^{-2x/\lambda_\perp}$, where $\lambda_\perp$ is the characteristic penetration length of the mode field amplitude into the surrounding vacuum (i.e., the impact-parameter-dependent excitation probability can be approximated as $dP_{\rm EELS}(x)/dz\approx\ee^{-2x/\lambda_\perp}\times[dP_{\rm EELS}(0)/dz]$). For example, one finds $\lambda_\perp=1/\sqrt{\kpar^2-\omega^2/c^2}$ for the fundamental band in a cylindrical waveguide with $m=0$ azimuthal symmetry, while we have $\lambda_\perp\approx150$~nm in the leftmost mode of the silicon waveguide considered in Fig.~\ref{Fig6}b (see supplementary Fig.~\ref{FigS5}).

In practice, one can use Gaussian e-beams running nearly parallel to the waveguide. For an e-beam with 1D Gaussian profile in the $xz$ plane and translational invariance along $y$, the electron probability density can be written as $|\psi(x,z)|^2=\big[(2/\pi)^{1/2}/w(z)\big]\,\ee^{-2[x/w(z)]^2}$ with width $w(z)=W_0\sqrt{1+(z\lambda_e/\pi W^2_0)^2}$ that evolves along the e-beam longitudinal direction $z$ as determined by the electron wavelength $\lambda_e$ (e.g., 2.5~pm at 200~keV) and the width $W_0$ at the waist ($\approx\lambda_e/\pi\,{\rm NA}$ for typical numerical apertures NA$\lesssim0.02$). The waveguide can be bent with a matching Gaussian curvature (i.e., a surface profile $x=-\xi w(z)$, which blocks a fraction $[1-{\rm erf}(\xi\sqrt{2})]/2$ of the electrons, for example, 1\% for $\xi=1.16$). This strategy should allow us to increase the interaction length, as illustrated in Fig.~\ref{Fig6}c. We now recall that the excitation probability is the inelastic average of the e-beam loss rate weighted with the electron density across the transverse profile \cite{RH1988,paper149} (see Eq.~(\ref{PpsiG})). For the sake of this discussion, the excitation probability can then be written as $\int dx\,|\psi(x,z)|^2\,[dP_{\rm EELS}(x)/dz]$. Assuming an adiabatic evolution of the e-beam profile along $z$ and noticing that the electron velocity is nearly parallel to the surface at all times and positions, the resulting excitation probability becomes $L_{\rm eff}\times[dP_{\rm EELS}(0)/dz]$, where $L_{\rm eff}=\int dz\int_0^\infty dx\,|\psi[x-\xi w(z),z]|^2\,\ee^{-2x/\lambda_\perp}$. For large $|z|\gg W_0^2/\lambda_e$, we have $w(z)\approx|z|\lambda_e/\pi W_0$, so the electron density scales as $|\psi[x-\xi w(z),z]|^2\propto1/|z|$ near the surface, and the $z$ integral diverges logarithmically with the length $\Delta z$ of the curved surface. This divergence indicates that $L_{\rm eff}$ can take large values, which should be physically limited by inelastic interactions and diffraction due to the presence of the surface. As an example involving conservative parameters, we take 200~keV electrons and $W_0=10$~nm, so that the $z$-dependent e-beam diameter becomes $w(z)=10\,{\rm nm}\times\sqrt{1+[z/(125\,\mu{\rm m})]^2}$, which, combined with $dP_{\rm EELS}(0)/dz\approx0.02/\mu$m for the silicon waveguide in Fig.~\ref{Fig6}b, should enable the excitation of many quanta per electron for an interaction length $L_{\rm eff}=\Delta z\sim200\,\mu$m over which the e-beam width $w(z)$ increases to $\sim20$~nm.

A simpler configuration consists in reflecting the electron with a small glancing angle $\theta_0\ll1$ on a straight-line waveguide (Fig.~\ref{Fig6}d). Assuming again an exponential decay of the mode outside the waveguide and considering an adiabatic evolution of the interaction with probability $dP_{\rm EELS}(x)/dz$ per unit of electron path length, as obtained for a classical point electron moving parallel to the waveguide, we calculate an effective interaction length $L_{\rm eff}\approx2\lambda_\perp/\theta_0$. For example, for $\theta_0=1$~mrad, assuming the above parameters for the leftmost silicon mode in Fig.~\ref{Fig6}b, we find $L_{\rm eff}\approx300\,\mu$m and $P_{\rm EELS}\approx 6$. For such low $\theta_0$, the out-of-plane electron kinetic energy is small compared with the potential inside the material, and thus, we anticipate nearly perfect electron reflection. Incidentally, the attraction exerted by image charges and currents should be taken into consideration for a reliable description of the electron trajectory, imposing a minimum $\theta_0$.

If the mode dispersion can be tailored such that it is tangent to the electron line $\omega=\kpar v$ (i.e., a mode dispersion containing one point in which the phase and group velocities are the same, and the electron velocity is tuned to match them), a stronger interaction is expected, changing the coupling scaling with the effective interaction length to a higher power of $L_{\rm eff}$ \cite{KRR24}. In general, we can consider a mode dispersion behavior $\omega_{\kpar}\approx\kpar v+\zeta(\kpar-k_c)^n$ near the matching point $\kpar=k_c$, which corresponds to a regular crossing with $\zeta=v_{k_c}-v$ for $n=1$, producing $P_{\rm EELS}\propto L_{\rm eff}$ (see above); an inflection point as considered in ref~\citenum{KRR24} for $n=2$, which leads to $P_{\rm EELS}\propto L_{\rm eff}^{3/2}$; or an even smoother tangent point for $n>2$. As shown in Methods, the coupling probability becomes
\begin{align}
P_{{\rm EELS}}=L_{\rm eff}^{2-1/n}\;\frac{S_n\alpha}{\beta k_c}\big(2v/\zeta\big)^{1/n}\,|E_{k_c,z}(\Rb_0)|^2
\label{lastlastlast}
\end{align}
with $S_n=(2/n)\int_0^\infty d\theta\,\theta^{1/n-3}\,\sin^2\theta$ (in particular, $S_1=\pi$ and, using eq~3.762-1 of ref~\citenum{GR1980}, $S_n=(2^{2-1/n}/n)\Gamma(1/n)\cos(\pi/2n)/[(1-1/n)(2-1/n)]$ for $n>1$). We have a scaling $P_{{\rm EELS}}\propto L_{\rm eff}^{2-1/n}$ that starts at $\propto L_{\rm eff}$ for $n=1$ and approaches $\propto L_{\rm eff}^2$ for high $n$.

These configurations for generating multiple photons per electron in a waveguide mode can be combined with light-optics schemes to couple them to localized excitations (Fig.~\ref{Fig6}e), such as those of single molecules \cite{WGH08,RWL12}. This approach could exceed the efficiency of direct e-beam coupling to localized modes discussed in previous sections as a route to deterministically entangle single photons and excitations in single molecules. Rather than relying on the direct interaction of the e-beam with a 3D-confined optical mode, one could instead generate multiple waveguided photons per electron, which are then funneled into localized excitations in an integrated-optics setup. We renounce leveraging the excellent spatial resolution of e-beams for mapping the targeted excitation modes, and instead, use such resolution to couple efficiently to waveguides and subsequently populate a localized excitation, whose correlation with the energy loss experienced by the electron should depend on the fidelity of the waveguide--localized-mode coupling scheme.

A relevant question concerns the degree of coherence of the generated waveguided photons, which has two interrelated aspects: the temporal coherence of different photons created by a single electron and the spectral coherence at the single-photon level. Temporal coherence requires the electron wavepacket to have a small duration compared with the optical period of the emitted light \cite{paper373}, and therefore, unless attosecond electron pulses are employed, different emitted photons have random relative phases. Spectral coherence refers to the phase associated with different frequency components of the emitted light, and whether we can identify individual photons consisting of the superposition of different colors with well-defined relative phases. For a given impact parameter (i.e., a specific electron--waveguide distance), the generated waveguide field contains a finite range of frequencies, with a spectral width $\Delta\omega$ that depends on the group velocity of the waveguide mode (i.e., the difference in slopes of the electron line and the mode dispersion, see Fig.~\ref{Fig6}a). This spectral width is acquired due to the finite time of electron--waveguide interaction and the spread in incident electron velocities among other factors. The frequency superposition can be regarded as coherent if $\hbar\Delta\omega$ is small compared with the energy spread of the incident electron wave function, as otherwise different spectral regions of the generated light would be associated with discernable energy losses in the electron, thus defining an incoherent set when tracing out the electron degrees of freedom. (Incidentally, the total electron spectral width is expected to be generally larger than this {\it coherent} spectral width, as it can also be contributed by incoherent components of the electron density matrix.) For example, considering an interaction length $L_{\rm eff}\sim10\,\mu$m (i.e., $\Delta t\sim48$~fs at 200~keV), the optical spectral width $\hbar/\Delta t\sim14$~meV already exceeds the zero-loss energy broadening of currently available monochromated e-beams \cite{HRK20,YLG21} ($\sim$ a few meV). This problem is alleviated when $L_{\rm eff}$ extends over hundreds of microns, as in the examples discussed above. In addition, different impact parameters produce incoherent superpositions of the generated optical fields \cite{paper373} (see Eq.~(\ref{PpsiG})), unless one post-selects (and heralds the emission) by detecting electrons that have lost which-way information \cite{paper388}. It should also be noted that these incoherent components have different spectral distributions because the excitation amplitude of each frequency component depends on the electron impact parameter \cite{paper149}, so they are not spectrally equivalent.

\begin{table*}
\centering
\caption{{\bf Overview of free-electron coupling to optical modes: scalings and magnitudes.} We consider different types of systems (leftmost column) characterized by a size $D$, focusing on one of their optical modes $i$ at frequency $\omega_i$. For excitation by an electron with velocity $v$, we define the phase $\varphi_i=\omega_iD/v$ (fourth column), which determines the coupling strength through a coupling function $F_i(\varphi_i)$ (the squared spatial Fourier transform of the mode field along the e-beam for a spatial frequency $\omega_i/v$). The table summarizes the scalings of the frequency-integrated EELS and CL probabilities $P_{{\rm EELS/CL},i}$ (second and third columns), as well as the maximum $P_{{\rm EELS},i}$ here obtained under practical configurations (rightmost column). The list of configurations includes (from top to bottom) a silver nanorod (length $D=20$~nm, $6.7$~nm diameter, Fig.~\ref{Fig1}e); the dipolar $m=0$ mode of a silver sphere (diameter $D=20$~nm, Fig.~\ref{Fig2}a);  the TM$_{201}$ mode in a metal-coated dielectric cylinder ($\epsilon=100$, length $D=1\,\mu$m, $40$~nm radius, Fig.~\ref{Fig3}b); the dipolar $m=0$ electric mode of dielectric Mie spheres (diameter $D=128$~nm with either $\epsilon= 32$ (absolute maximum in supplementary Fig.~\ref{FigS3}a) or $\epsilon=15$); the $m=1$ mode in 2D graphene (diameter $D=18$~nm, 0.4~eV Fermi energy, Fig.~\ref{Fig4}c) and monolayer WS$_{2}$ (diameter $D=10$~nm, Fig.~\ref{Fig4}e) disks; and hydrogen-like s-p transitions (orbital size $D$, p-orbital orientation parallel or perpendicular to e-beam, Fig.~\ref{Fig5}d,e). The bottom row refers to the excitation of 1D waveguide modes exemplified by the rectangular silicon rod in Fig.~\ref{Fig6}b (width $D=400$~nm) with an effective interaction length $L_{\rm eff}=300\,\mu$m under the configuration of Fig.~\ref{Fig6}d for a glancing angle $\theta_0=1$~mrad.}
\includegraphics[width=1.0\textwidth]{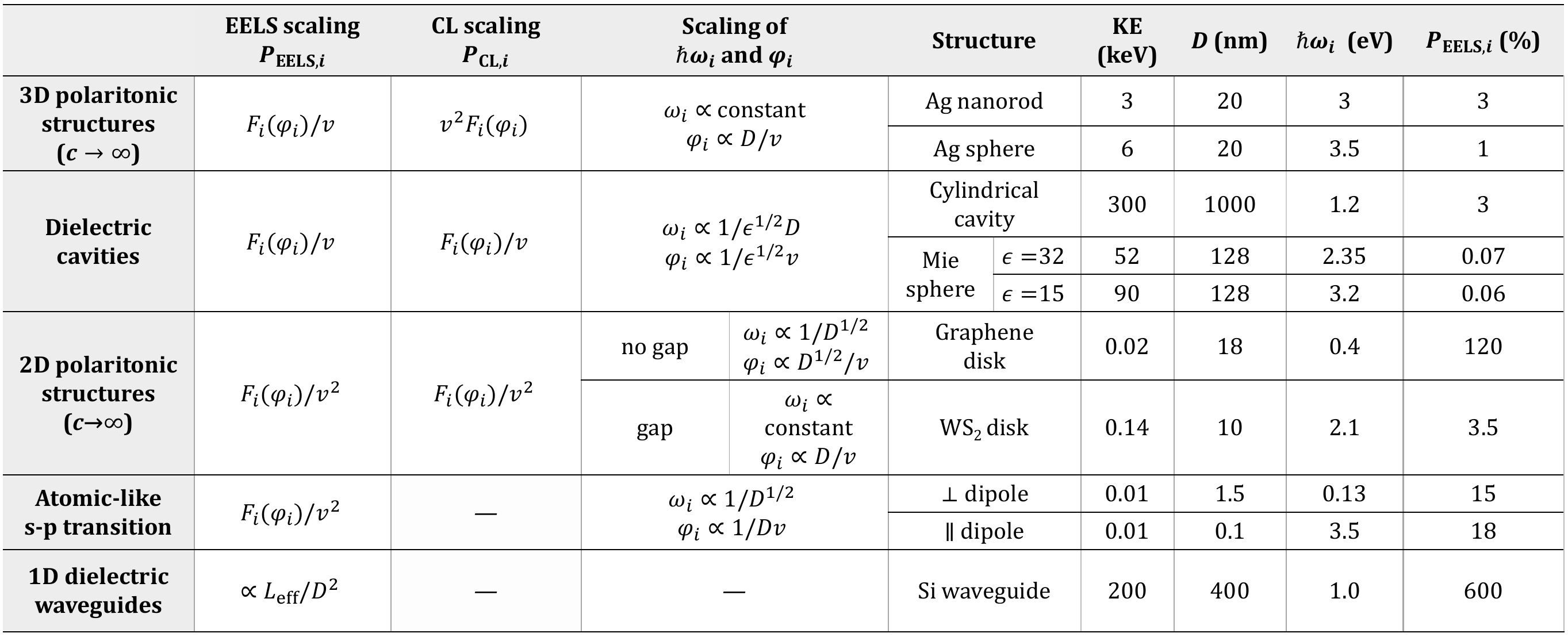}
\label{Table2}
\end{table*}

\section{Conclusions}

In conclusion, by formulating simple scaling laws for the EELS and CL probabilities as a function of the size of the specimen and the electron velocity, we perform an extensive exploration of the inelastic coupling between free electrons and confined optical modes, finding the conditions under which unity-order excitation probabilities are possible and establishing practical limits to the performance of general kinds of systems. Specifically, we consider dielectric cavities, small 2D and 3D polaritonic structures, atomic-like transitions, and waveguide modes. A summary of the results is presented in Table~\ref{Table2}, along with the scaling properties for the excitation probability and the mode frequencies. For a given mode $i$ of frequency $\omega_i$, the excitation probability is proportional to the squared spatial Fourier transform of the mode field along the e-beam direction ($F_i(\varphi_i)$, Eq.~(\ref{Fi})) evaluated at a spatial frequency $\omega_i/v$. This defines a phase $\varphi_i=\omega_iD/v$ when scaled by a characteristic distance $D$ of the system. A maximum excitation probability takes place when $\varphi_i\sim\pi$, with the actual optimum value depending on the specific material and morphology under consideration.

For polaritonic structures in the quasistatic regime ($\omega_iD/c\ll1$), $\omega_i$ remains approximately independent of size and only varies with shape. We then find a monotonic growth of the excitation probability as observed in EELS with decreasing $D$, and accordingly, the velocity needs to also be reduced to maintain the optimum value of $\varphi_i$. For the 3D geometries summarized in Table~\ref{Table2}, the coupling probabilities can exceed 1\% for structure sizes of tens of nm interacting with electrons in the keV range. We find the largest interaction probabilities (scaling as $\propto1/\beta$) under mode-phase-matching conditions in materials without an intrinsic resonance (in contrast to $\propto1/\beta^2$ when $\omega_g\neq 0$), such as 2D graphene nanoislands, for which values exceeding $100$\% are obtained with electron energies of tens of eV and diameters of tens of nm (see Table~\ref{Table2}) \cite{paper228}.

Importantly, as a consequence of the two-step process underlying CL emission (i.e., excitation by the electron and decay into radiation), the probability is proportional to $\beta^2$, thus suppressing the benefits that we find in EELS when coupling to low-energy electrons and setting the optimal parameters in a region where retardation effects may play a leading role. This conclusion is supported by an experimental study published together with the present work \cite{paperx2}, in which the CL intensity measured from gold nanospheres is shown to reach maximum values for optimum electron energies as indicated in Eq.~(\ref{quasicldep}), also showing that larger particles produce stronger signals.

We also discuss lossless dielectric structures, which, in contrast to polaritonic particles, display a scaling $\omega_i \propto 1/D$, so small mode frequencies $\omega_i$ are not compatible with small sizes unless unrealistically large permittivities are considered. A high excitation probability then requires long interaction lengths \cite{paper180,K19}. In this respect, it has been argued \cite{K19} that, for dielectric waveguides extending over a longitudinal distance $L_{\rm eff}$, the normalized mode electric field scales as $1/\sqrt{L_{\rm eff}}$, while, under phase-matching conditions, the coupling amplitude receives an additional factor of $L_{\rm eff}$, such that $F_i(\varphi_i)\propto L_{\rm eff}$. Although one could in principle achieve arbitrarily large excitation probabilities by elongating the waveguide, the e-beam is required to be laterally confined along an increasingly long propagation distance $L_{\rm eff}$, which imposes a constrain on the beam-structure distance due to diffraction-induced divergence along the propagation direction. In this work, we analyze practical configurations for coupling to waveguides, specifically considering either Gaussian e-beams running close to curved waveguides or plane-wave e-beams under glancing incidence on 1D straight-line waveguides. We find that several photons can be generated per electron under phase-matching conditions, thus establishing an alternative strategy to couple single electrons and single localized optical excitations efficiently by first generating waveguided photons that are subsequently coupled to localized excitations through light-optics schemes (see Fig.~\ref{Fig6}).

In electron-driven atomic-like transitions, we find the opposite dependence on the size of the system relative to 2D and 3D polaritonic structures: the interaction probability grows with size due to the $1/D^2$ scaling of the excitation frequency. We illustrate this effect for hydrogenic s~$\to$~p orbital transitions (Table~\ref{Table2}). The orbital size can be controlled in practice in artificial systems such as defects in TMDs \cite{paper354} and quantum dots\cite{BCK99}. Excitation probabilities nearing 20\% are predicted for 10~eV electrons, while higher values are possible due to the $1/v^2$ scaling of the excitation probability. Achieving such a high coupling between free electrons and two-level systems is important to trigger and explore nonlinear optical responses at the few-photon level, but also to develop a novel platform for quantum-optics protocols based on nanoscale systems in which, for example, an electron can probe quantum blockade or materialize delayed-choice experiments \cite{MKZ16}.

The present results pave the way toward a complete understanding of the inelastic coupling between free electrons and confined optical excitations, establishing general rules to maximize the interaction and providing scaling laws that involve the size of the physical system and the electron velocity. The list of unresolved problems that could benefit from the present study include the use of free electrons to perform spectroscopy on individual molecules, possibly coupled to cavities that mediate an optimum interaction to free electrons; the creation of multiple excitations by the same electron to probe the nonlinear dynamics of nanostructures and materials in general, such as, for example, quantum-optics effects in the few-body limit; the generation of high-order Fock states $\ket{n}$ in an optical mode of frequency $\omega$ triggered by a more efficient probability of finding electrons that have lost an energy $n\hbar\omega$; and the extraction of more information per electron when performing spectrally resolved microscopy. Our work may lay down a way to tackle these research challenges from a fresh perspective, invoking the strong electron--light interaction regime to grant us access to unexplored fundamental phenomena such as the generation of nonclassical light and the complex nonlinear processes taking place in out-of-equilibrium nanostructures down to atomic scales.

\begin{widetext}

\section*{Appendix}
\appendix

\section{Electrostatic Mode Expansion in 3D Structures}
\renewcommand{\theequation}{A\arabic{equation}}
\label{AppendixA}

We briefly describe a well-known method used to solve the quasistatic problem in terms of a mode expansion of the surface charges induced by an external perturbation (see, for example, refs~\citenum{OI1989,paper010,BK12}). Working in frequency space $\omega$, we begin by writing the Poisson equation corresponding to the total scalar potential $\phi(\rb,\omega)=\int dt\,\ee^{\ii\omega t}\phi(\rb,t)$ created by an external charge density $\rho^{\rm ext}(\rb,\omega)$ in the presence of a structure described by a complex, frequency-dependent dielectric function $\epsilon(\rb,\omega)$:
\begin{align}
\nabla\cdot[\epsilon(\rb,\omega) \nabla\phi(\rb,\omega)]=-4\pi \rho^{\rm ext}(\rb,\omega)\label{pois}.
\end{align}
A formal solution of Eq.~(\ref{pois}) can be written as the sum of {\it external} and {\it induced} potentials,
\begin{subequations}
\begin{align}
\phi^{\rm ext}(\rb,\omega)&=\int d^3\rb' \;\frac{1}{|\rb-\rb'|}\;\frac{\rho^{\rm ext}(\rb',\omega)}{\epsilon(\rb',\omega)},\label{pext}\\
\phi^{\rm ind}(\rb,\omega)&=\int d^3 \rb' \,\frac{\Db(\rb',\omega)\cdot \nabla_{\rb '} [1/\epsilon(\rb',\omega)]}{4\pi |\rb-\rb'|},\label{pind}
\end{align}
\end{subequations}
with $\Db(\rb,\omega)=-\epsilon(\rb,\omega) \nabla_\rb \phi(\rb,\omega)$. We focus on systems involving two values of the dielectric function, $\epsilon(\omega)$ and $\epsilon_h(\omega)$, in the regions occupied by a homogeneous nanostructure (defined by a step function $\Theta(\rb)=1$) and a host medium ($\Theta(\rb)=0$), respectively, separated by an abrupt interface $S$. This allows us to write $\nabla_\rb[1/\epsilon(\rb,\omega)]=(1/\epsilon_h-1/\epsilon)\,\nn_\sb\,\delta_\sb$, where $\delta_\sb$ is a surface $\delta$-function that restricts $\rb$ to interface points $\sb$, while $\nn_\sb$ is the unit vector perpendicular to the interface at $\sb$ and pointing towards the host medium. Then, we can rewrite Eq.~(\ref{pind}) in terms of the surface-charge density $\sigma(\sb,\omega)=[(\epsilon-\epsilon_h)/4\pi \epsilon\epsilon_h]\,\nn_\sb\cdot \Db(\sb,\omega)$ as
\begin{align}
\phi^{\rm ind}(\rb,\omega)=\oint_S d^2\sb \, \frac{\sigma(\sb,\omega)}{|\rb-\sb|}.
\end{align}
Assuming $\rho^{\rm ext}(\rb,\omega)$ to be free of singularities on $S$, the continuity of the normal displacement $\nn_\sb\cdot\Db(\sb,\omega)$ at the interface allows us to write \cite{paper010}
\begin{align}
\nn_\sb\cdot \Db(\sb,\omega)
&=\epsilon_h \left[-\nn_\sb\cdot\nabla_\sb \phi^{\rm ext}(\sb,\omega)-\oint_S d^2\sb' F(\sb,\sb')\sigma(\sb',\omega)+2\pi \sigma(\sb,\omega)\right] \nonumber\\
&=\epsilon \left[-\nn_\sb\cdot\nabla_\sb \phi^{\rm ext}(\sb,\omega)-\oint_S d^2\sb' F(\sb,\sb')\sigma(\sb',\omega)-2\pi \sigma(\sb,\omega)\right], \label{nsD}
\end{align}
where we introduce the continuous part of the normal derivative of the Coulomb field  $F(\sb,\sb')=-\nn_\sb \cdot (\sb-\sb')/|\sb-\sb'|^3$ and separate the singular part at $\sb'\rightarrow\sb$ giving rise to terms proportional to $2\pi$ \cite{paper010}. From Eq.~(\ref{nsD}), we obtain the self-consistent equation
\begin{align}
2\pi\lambda(\omega)\sigma(\sb,\omega)=\nn_\sb\cdot \nabla_\sb \phi^{\rm ext}(\sb,\omega)+\oint_S d\sb' \, F(\sb,\sb') \sigma(\sb',\omega)
\label{selfconsistentsigma}
\end{align} 
for the charge density, where $\lambda(\omega)=(\epsilon_h+\epsilon)/(\epsilon_h-\epsilon)$. The operator $\hat{O}$ defined by $[\hat{O}\cdot f](\sb)=\oint_S d\sb' F(\sb,\sb')f(\sb')$ can be proven to admit a complete set of real eigenvalues $\lambda_i$ and eigenfunctions $\sigma_i(\sb)$ such that \cite{OI1989}
\begin{align}
\oint_S d\sb' F(\sb,\sb') \sigma_i(\sb')=2\pi \lambda_i\sigma_i(\sb).\label{eigen}
\end{align}
Moreover, the eigenfunctions fulfill the orthogonality relation \cite{OI1989}
\begin{align}
\oint_S d^2\sb \oint_S d^2\sb' \;\frac{\sigma_i(\sb)\sigma_{i'}(\sb')}{|\sb-\sb'|}=\frac{\delta_{ii'}}{D}, \label{norm}
\end{align}
where we introduce a characteristic distance of the structure $D$ for normalization. We now expand the surface charge density as a linear superposition of $\sigma_i(\sb)$ terms. The expansion coefficients are then obtained from Eqs.~(\ref{selfconsistentsigma}) and (\ref{norm}), leading to
\begin{subequations}
\label{essolutions}
\begin{align}
\sigma(\sb,\omega)&=\sum_i g_i(\omega) f_i(\omega) \sigma_i(\sb),\\
g_i(\omega)&=\frac{1}{\lambda(\omega)-\lambda_i}=\frac{\epsilon_h-\epsilon}{(1-\lambda_i)\epsilon_h+(1+\lambda_i)\epsilon}, \label{gi}\\
f_i(\omega)&=\frac{D}{2\pi} \oint_S d^2\sb\oint_S d^2\sb' \;\frac{\sigma_i(\sb')}{|\sb-\sb'|}
\;\nn_\sb\cdot \nabla_\sb \phi^{\rm ext}(\sb,\omega) \label{essolutions3}\\
&=D \int d^3\rb\;\phi_i(\rb)\;\frac{\rho^{\rm ext}(\rb,\omega)}{\epsilon(\rb,\omega)}\;\big[\lambda_i+1-2\Theta(\rb)\big], \label{essolutions4}
\end{align}
\end{subequations}
where Eq.~(\ref{essolutions4}) involves the eigenpotential
\begin{align}
\phi_i(\rb)=\oint_S d^2\sb \,\frac{\sigma_i(\sb)}{|\rb-\sb|}. \label{phii}
\end{align}
Note that Eq.~(\ref{essolutions4}) is obtained from Eq.~(\ref{essolutions3}) by applying Eq.~(\ref{eigen}) after using the properties of $\int_S d^2\sb \, F(\sb,\rb)/|\sb-\rb'|$ following from Green's identity upon exchange of $\rb$ and $\rb'$  when these points lie exactly at surface, on opposite sides of the interface, or the same side \cite{OI1989}. From here, we write the solution of Eq.~(\ref{pois}) as
\begin{align}
\phi(\rb,\omega)=\int d^3\rb'\; \mathcal{W}(\rb,\rb',\omega)\;\rho^{\rm ext}(\rb',\omega),
\label{phisolution}
\end{align}
where
\begin{align}
\mathcal{W}(\rb,\rb',\omega)&=\frac{1}{\epsilon(\rb',\omega)}\left[\frac{1}{|\rb-\rb'|}+D\sum_i \big[\lambda_i+1-2\Theta(\rb')\big]\,g_i(\omega) \phi_i(\rb)\phi_i(\rb')\right] \label{3dW}
\end{align}
is the screened interaction $\mathcal{W}(\rb,\rb',\omega)$, which enters naturally the calculation of EELS and CL probabilities (see below). We interpret $\mathcal{W}(\rb,\rb',\omega)$ as the potential created at $\rb$ by a unit charge placed at $\rb'$ and oscillating with frequency $\omega$. Incidentally, reciprocity requires the symmetry property $\mathcal{W}(\rb,\rb',\omega)=\mathcal{W}(\rb',\rb,\omega)$, which implies the completeness relation $D\sum_i\phi_i(\rb)\phi_i(\rb')=1/|\rb-\rb'|$ for $\rb$ and $\rb'$ on opposite sides of the interface.

The absence of an absolute length scale in the quasistatic approximation is reflected in the fact that the eigenvalues $\lambda_i$ are independent of the size of the system. However, according to Eq.~(\ref{norm}), the eigenfunctions change with the size of the system $D$ according to the rule $\sigma_i(\sb)=D^{-2}\tilde{\sigma}_i(\sb/D)$, where $\tilde{\sigma}_i$ are scale-invariant functions. This implies the property $\phi_i(\rb)=D^{-1}\tilde{\phi}_i(\rb/D)$, where $\tilde{\phi}_i(\ub)$ is also a scale-invariant function of the dimensionless coordinates $\ub=\rb/D$.

\section{Electrostatic Mode Expansion in 2D Structures}
\renewcommand{\theequation}{B\arabic{equation}}
\label{AppendixB}

We consider a planar 2D structure placed in the $z=0$ plane, embedded in a host medium of permittivity $\epsilon_h$, and having a small thickness compared to both the lateral size and the characteristic distances over which the optical field varies significantly. This type of system has been analyzed through quasistatic modal expansions with the material assimulated to a zero-thickness film of frequency-dependent surface conductivity $\sigma(\omega)$ \cite{paper228}. Such expansions can be constructed from the self-consistently induced surface charge \cite{paper228}, but they can also be derived from the formalism presented above for 3D structures by taking the limit of a vanishing thickness $d\to0^+$ combined with a permittivity $\epsilon(\omega)=4\pi\ii\sigma(\omega)/\omega d$ that embodies the finite surface conductivity. For example, we can rewrite Eq.~(\ref{3dW}) for $\rb$ and $\rb'$ in the host medium as
\begin{align}
\mathcal{W}(\rb,\rb',\omega)=\frac{1}{\epsilon_h}\left[\frac{1}{|\rb-\rb'|}+D\sum_i\frac{\phi_i(\rb) \phi_i (\rb')}{\eta_i/\eta(\omega)-1} \right], \label{2dW}
\end{align}
where we have made use of Eq.~(\ref{gi}), neglected $\epsilon_h$ compared with $\epsilon$, and defined the parameters $\eta(\omega)=\ii\sigma(\omega)/\epsilon_h\omega D$ and $\eta_i=(d/4\pi D)[(\lambda_i-1)/(\lambda_i+1)]$. We recall that this analysis is based on the continuity of the normal displacement at the material interface (i.e., Eq.~(\ref{selfconsistentsigma})). An equivalent procedure for zero-thickness structures consists of writing a self-consistent equation for the in-plane electric field \cite{paper228}. Although the link between these two approaches is rather involved, the latter leads to the same expression as in Eq.~(\ref{2dW}), where each $\eta_i$ is found to take a finite value in the $d\to0^+$ limit, arising from an eigenvalue $\lambda_i\to-1$. In addition, $\eta_i$ is negative and independent of $D$, while $\phi_i(\rb)$ satisfies the same scaling property as deduced above for 3D structures. 

\section{Scaling of the EELS and CL Probabilities in 3D Dielectric Cavities}
\renewcommand{\theequation}{C\arabic{equation}}
\label{AppendixC}

We are interested in deriving general scaling properties of the EELS and CL probabilities using Eqs.~(\ref{eels}) and (\ref{cl}) for different types of cavities. We start by considering dielectric cavities with the inclusion of retardation effects. For a structure made of a self-standing lossless material characterized by a real, frequency-independent permittivity $\epsilon$, Eq.~(\ref{gt}) readily leads to the following property for the Green tensor \cite{BSB10}:
\begin{align}
\tilde{G}(\rb,\rb',\omega)=\frac{1}{\mu}\,G(\rb/\mu,\rb'/\mu,\mu\omega), \quad\quad\text{[dielectrics]} \label{gtscaldiel}
\end{align} 
where the tilde indicates that we refer to a system with the same morphology but in which all distances are scaled by a factor $\mu$ (e.g., $\mu>1$ for expansion). Using this result in combination with Eqs.~(\ref{eels}) and (\ref{cl}), we obtain
\begin{subequations}
\begin{align}
\tilde{\Gamma}_{\rm EELS}(\Rb_0,v,\omega)&=\mu\Gamma_{\rm EELS}(\Rb_0/\mu,v,\mu\omega),\nonumber\\
\frac{d\tilde{\Gamma}_{\rm CL}(\Rb_0,v,\omega)}{d\omega d\Omega_{\rr}}&=\mu\frac{d\Gamma_{\rm CL}(\Rb_0/\mu,v,\mu\omega)}{d\omega d\Omega_{\rr}}.\nonumber
\end{align} 
\end{subequations}
The spectral signature associated with a mode resonance $i$ can be integrated over frequency by using these expressions to obtain the following scaling property for the mode excitation probabilities: $\tilde{P}_{{\rm EELS/CL},i}(\Rb_0,v)=P_{{\rm EELS/CL},i}(\Rb_0/\mu,v)$. We recall that the EELS and CL probabilities must be identical under the assumption of lossless dielectrics, and therefore, $P_{{\rm EELS},i}(\Rb_0,v)=P_{{\rm CL},i}(\Rb_0,v)$.

To derive the functional dependence of Eq.~(\ref{Peelscldiel}), we expand the transverse part of the Green tensor in terms of normal modes\cite{GL91} 
\begin{align}
G(\rb,\rb',\omega)=\sum_i\frac{\Eb_i(\rb)\otimes \Eb_i^*(\rb')}{\omega^2-\omega_i^2+\ii0^+}, \label{GreenEiEi}
\end{align}
where $\Eb_i(\rb)$ are mode fields satisfying the wave equation $\nabla\times\nabla\times\Eb_i(\rb)=(\omega_i/c)^2\epsilon(\rb)\Eb_i(\rb)$ and the orthogonality relation $\int d^3\rb \,\epsilon(\rb) \, \Eb_i(\rb)\cdot \Eb_{i'}^*(\rb)=\delta_{ii'}$ with $\epsilon(\rb)=1+\Theta(\rb)(\epsilon-1)$. Incidentally, Eq.~(\ref{gtscaldiel}) is readily verified by noticing the scaling relations $\tilde{\Eb}_i(\rb)=\Eb_i(\rb/\mu)/\mu^{3/2}$ and $\tilde{\omega}_i=\omega_i/\mu$ imposed by  normalization and the wave equation. Now, inserting $G(\rb,\rb',\omega)$ into the EELS probability (Eq.~(\ref{eels})), using the noted scaling relations, integrating over frequency around the mode spectral width, and considering a small radiative mode damping rate ($\ll\omega_i$), we obtain Eq.~(\ref{Peelscldiel}) in the main text.

\section{Scaling of the EELS Probability in the Quasistatic Limit: 3D Structures}
\renewcommand{\theequation}{D\arabic{equation}}
\label{AppendixD}

For small structures compared to the wavelengths $\lambda_i=2\pi c/\omega_i$ associated with the excitation frequencies $\omega_i$, we can work in the quasistatic limit ($c\rightarrow \infty$) and calculate the EELS probability using Eq.~(\ref{eelsquasi}). Upon inspection of Eq.~(\ref{3dW}), using the scaling properties of $\phi_i(\rb)$ (see above), we find
\begin{align}
\tilde{\mathcal{W}}(\rb,\rb',\omega)=\frac{1}{\mu}\,\mathcal{W}(\rb/\mu,\rb'/\mu,\omega),\label{wscal}
\end{align}
which allows us to write the scaling law 
\begin{align}
\tilde{\Gamma}_{\rm EELS}(\Rb_0,v,\omega)=\frac{1}{\mu}\,\Gamma_{\rm EELS}(\Rb_0/\mu,v/\mu,\omega),  \label{quasieelsscal}
\end{align}
where again, quantities with tilde refer to a system in which distances are multiplied by a factor $\mu$. By integrating over frequency, the mode excitation probability scales as $\tilde{P}_{{\rm EELS},i}(\Rb_0,v)=\mu^{-1}P_{{\rm EELS},i}(\Rb_0/\mu,v/\mu)$ (Eq.~(\ref{Peelsscaling1})).

When considering aloof trajectories with the electron moving always in vacuum ($\epsilon_{\rm h}=1$), Eqs.~(\ref{3dW}) and (\ref{eelsquasi}) lead to
\begin{align}
\Gamma_{\rm EELS}(\Rb_0,v,\omega)=\frac{e^2D} {\pi \hbar v^2}\sum_i {\rm Im}\{-(\lambda_i+1)g_i(\omega)\}\left|\int_{-\infty}^\infty du_z\;\tilde{\phi}_i(\Rb_0/D,u_z)\ee^{-\ii \frac{\omega D}{v}u_z}\right|^2\label{tmp1},
\end{align}
 where $u_z=z/D$. Now, we assume a weakly absorbing medium (${\rm Im}\{\epsilon\}\ll|{\rm Re}\{\epsilon\}|$) and approximate the loss function (see Eq.~(\ref{gi})) as
\begin{align}
{\rm Im}\{-(\lambda_i+1)g_i(\omega)\}\approx ({\rm Re}\{\epsilon(\omega_i)\}-1)\,{\rm Im}\Bigg\{\frac{1}{{\rm Re}\{\partial_{\omega_i}\epsilon(\omega_i)\}(\omega-\omega_i)+i\,{\rm Im}\{\epsilon(\omega_i)\}}\Bigg\}. \nonumber
\end{align}
This expression is obtained by setting the real mode frequency $\omega_i$ such that ${\rm Re}\{\epsilon(\omega_i)\}=(\lambda_i-1)/(\lambda_i+1)$ (i.e., from the corresponding zero of the real part of the denominator in $g_i$), linearizing the frequency dependence of $\epsilon(\omega)$, and neglecting ${\rm Im}\{\partial_{\omega_i}\epsilon(\omega_i)\}$. Then, integrating over $\omega$ and assuming that the integral factor in Eq.~(\ref{tmp1}) does not vary significantly within the width of the resonance, we obtain
\begin{align}
P_{{\rm EELS},i}(\Rb_0,v)\approx\frac{\alpha}{\beta}\left(\frac{1-{\rm Re}\{\epsilon(\omega_i)\}}{\omega_i |{\rm Re}\{\partial_\omega \epsilon(\omega_i)\}|}\right) F_i\left(\varphi_i\right)
\end{align}
(i.e., Eq.~(\ref{quasieelsdep})), where $F_i\left(\varphi_i\right)$ is given by Eq.~(\ref{Fi}). To derive this result, we integrate by parts in Eq.~(\ref{tmp1}) and write the scaled electric field as $\tilde{E}_{i,z}=-\partial_{u_z}\tilde{\phi}_i$ in terms of the potential. Importantly, the probability is guaranteed to be positive by the fact that $|\lambda_i|<1$ \cite{OI1989}.

\section{Scaling of the CL Probability in the Quasistatic Limit: 3D Structures}
\renewcommand{\theequation}{E\arabic{equation}}
\label{AppendixE}

In the quasistatic limit, the screened interaction $\mathcal{W}(\rb,\rb',\omega)$ accurately approximates the longitudinal component of the near field close to the sample, but to compute CL, we need to evaluate the far-field scattering amplitude $\fb_\rr(\rb,\omega)$, which involves the transverse component of the Green tensor in the $kr\rightarrow \infty$ limit (see Eq.~(\ref{cl})), with $k=\omega/c$. To this end, we use the Dyson equation associated with $G(\rb,\rb',\omega)$,
\begin{align}
G(\rb,\rb',\omega)=G_0(\rb,\rb',\omega)+\omega^2 (1-\epsilon)\int_V d^3\rb''\, G_0(\rb,\rb'',\omega)G(\rb'',\rb',\omega),\label{dys}
\end{align}
where the integral is restricted to the volume $V$ occupied by the dielectric and we introduce the free-space Green tensor\cite{J99} $G_0(\rb,\rb',\omega)=(-1/4\pi\omega^2)(k^2\mathcal{I}+\nabla_\rb \otimes \nabla_\rb)\ee^{\ii k|\rb-\rb'|}/|\rb-\rb'|$ with $\mathcal{I}$ standing for the 3$\times$3 identity matrix. By taking the far-field limit $\lim\limits_{kr \to \infty}G_0(\rb,\rb',\omega)=(1/4\pi c^2)\ee^{\ii k (r-\rr\cdot \rb')}(\rr\otimes \rr-\mathcal{I})/r$, approximating the right-most Green tensor in Eq.~(\ref{dys}) as $G(\rb,\rb',\omega)\approx \nabla_\rb \otimes \nabla_{\rb'} \mathcal{W}(\rb,\rb',\omega)$ $/4\pi\omega^2$ in terms of the screened interaction, and integrating by parts, we obtain the far-field amplitude
\begin{align}
\fb_\rr(\Rb_0,\omega)=&\frac{ek^2}{4\pi v}(1-\epsilon) \int_Vd^3\rb''\, \ee^{-\ii k \rr\cdot \rb''} \nonumber\\
&\times[\rr(\rr\cdot \nabla_{\rr''})-\nabla_{\rb''}]\int_{-\infty}^\infty dz' \mathcal{W}(\rb'',\Rb_0,z',\omega)  \ee^{\ii \omega z'/v}.\label{ffquasi}
\end{align}
To relate the CL emission probability $d\tilde{\Gamma}_{\rm CL}(\Rb_0,v,\omega)/d\omega$ in a system in which all distances have been scaled by a factor $\mu$ to the probability $d\Gamma_{\rm CL}(\Rb_0,v,\omega)/d\omega$ in the original system, we substitute $\mathcal{W}(\rb'',\Rb_0,z',\omega)$ by $\tilde{\mathcal{W}}(\rb'',\Rb_0,z',\omega)$ in Eq.~(\ref{ffquasi}), integrate over $\rb''$ within the scaled volume, and use the scaling law in Eq.~(\ref{wscal}). In addition, we Taylor-expand the exponential $\ee^{-\ii k \rr\cdot \rb''}$ inside the $\rb''$ integrand and write $\tilde{\fb}_\rr(\Rb_0,\omega)=\sum_{n=0}^\infty \alpha_n  \mu^{n+1} \fb_{\rr}^{(n)}(\Rb_0,\omega)$ with $\alpha_n=(n!)^{-1}$ and
\begin{align}
\fb_{\rr}^{(n)}(\Rb_0,\omega)=&\frac{ek^2\mu}{4\pi v}(1-\epsilon) \int_{V} d^3\wb \;(-\ii k\,\wb\cdot\rr)^n \nonumber\\
&\times[\rr(\rr\cdot \nabla_{\wb})-\nabla_{\wb}]\int_{-\infty}^\infty dw'_z \,\mathcal{W}(\wb,\Rb_0/\mu,w_z',\omega)
\ee^{\ii (\omega \mu/v) w_z'}, \nonumber
\end{align}
where the variable of integration $\wb=\rb''/\mu$ is normalized using the scaling factor $\mu$. Finally, we insert the far-field amplitude into Eq.~(\ref{cl}) to derive the sought-after CL scaling
\begin{align}
\frac{d\tilde{\Gamma}_{\rm CL}(\Rb_0,v,\omega)}{d\omega }=\sum_{n=0}^\infty \mu^{n+2}\,\frac{d\Gamma_{{\rm CL}}^{(n)}(\Rb_0/\mu,v/\mu,\omega)}{d\omega},\label{ffarl}
\end{align}
where 
\begin{align}
\frac{d\Gamma_{{\rm CL}}^{(n)}(\Rb_0/\mu,v/\mu,\omega)}{d\omega} =\sum_{m=0}^n\frac{\alpha_m \alpha_{n-m}}{4\pi^2 k\hbar}\int d\Omega_{\rr} \,\fb_{\rr}^{(m)}(\Rb_0,\omega)\fb_{\rr}^{(n-m)*}(\Rb_0,\omega). \nonumber
\end{align}
By integrating Eq.~(\ref{ffarl}) over the mode spectral width and defining
\[
P_{{\rm CL},i}^{(n)}(\Rb_0,v)=\int_i d\omega \, [d\Gamma_{{\rm CL}}^{(n)}(\Rb_0,v,\omega)/d\omega],
\]
we obtain the scaling relation in Eq.~(\ref{Pclscaling}).

Because the CL probability is the square of the far-field amplitude, it contains contributions arising from the interference between different modes \cite{LK15}. For simplicity, we focus here on systems in which such modes are spectrally separated and mode interference can be disregarded. By again taking the electron to follow a trajectory entirely contained in vacuum, we insert Eq.~(\ref{3dW}) into Eq.~(\ref{cl}) and use the scaling properties of the eigenpotentials to obtain the CL probability integrated over all solid angles:
\begin{align}
\frac{d\Gamma_{{\rm CL}}(\Rb_0,v,\omega)}{d\omega}\approx \frac{e^2 \omega^3 D^4}{64 \pi^4 \hbar c^3 v^2}& \left|\frac{(1-\epsilon)^2(1+\lambda_i)}{\epsilon(1+\lambda_i)+(1-\lambda_i)}\right|^2 \left| \int_{-\infty}^\infty du_z \tilde{\phi}_i(\Rb_0/D,u_z)\ee^{-\ii \frac{\omega D}{v} u_z}\right|^2 \label{tmp2}\\
\times &\int d\Omega_\rr \left|\int d^3\ub\,\ee^{-\ii \omega D (\rr\cdot\ub)/c}\left[\nabla_\ub\tilde{\phi}_i(\ub)-\rr (\rr\cdot\nabla_\ub\tilde{\phi}_i(\ub))\right]\right|^2, \nonumber
\end{align}
where $\ub=\rb/D$ is the position coordinate normalized to the characteristic distance $D$ in the structure. We work in the small-damping limit, by analogy to the discussion of EELS above. Then, integrating over frequency, we obtain
\begin{align}
P_{{\rm CL},i}(\Rb_0,v)\approx \beta^{2} \alpha  \left[\frac{(1-{\rm Re}\{\epsilon(\omega_i)\})^4}{64 \pi^3\omega_i {\rm Re}\{\partial_{\omega_i} \epsilon(\omega_i)\}{\rm Im}\{\epsilon(\omega_i)\}} \right]\chi_i(\beta,\varphi_i) \;\varphi_i^3F_i\left(\varphi_i\right), \label{pclquas}
\end{align}
where we introduce the radiating function
\begin{align}
\chi_i(\beta,\varphi_i)=\int d\Omega_\rr \left|\int d^3\ub\,\ee^{-\ii \varphi_i \beta (\rr\cdot\ub)}\left[\nabla_\ub\tilde{\phi}_i(\ub)-\rr (\rr\cdot\nabla_\ub\tilde{\phi}_i)(\ub)\right]\right|^2,
\label{chii}
\end{align}
and use $F_i(\varphi_i)$ as given in Eq.~(\ref{Fi}). Finally, we absorb some of the factors of Eq.~(\ref{pclquas}) into a mode-dependent constant $B_i$ to write the CL scaling law in Eq.~(\ref{quasicldep}).

\section{Scaling of the EELS and CL Probabilities in the Quasistatic Limit: 2D Structures}
\renewcommand{\theequation}{F\arabic{equation}}
\label{AppendixF}

For flat samples of vanishing thickness, we follow similar steps as for 3D systems, but with the screened interaction replaced by Eq.~(\ref{2dW}) and using the surface conductivity in Eq.~(\ref{sigmaD}). We focus on modes $i$ that have large frequencies $\omega_i$ compared with the damping rate $\gamma$. From the corresponding poles in Eq.~(\ref{2dW}) (i.e., $\eta(\omega_i)=\eta_i$), we find $\omega_i=\omega_g/2+\sqrt{\omega_g^2/4-e^2\omega_D/D\hbar\eta_i}$ (Eq.~(\ref{wi2D})), where $\omega_g$ and $\omega_D$ are inherited from Eq.~(\ref{sigmaD}) and we have neglected $\gamma$. After some algebra, the frequency-integrated EELS probabilities for exciting the modes under consideration are found to satisfy Eq.~(\ref{2dpeels}), whereas the CL probabilities are given by
\begin{align}
P_{{\rm CL},i}(\Rb_0,v)\approx \frac{\alpha^{5}}{8\pi \beta^2} \frac{\omega_D^4}{\eta_i^2\gamma\omega_i[(\omega_i-\omega_g)^2+\gamma^2]}  \chi_i(\beta,\varphi_i) \,\frac{1}{\varphi_i}F_i\left(\varphi_i\right)\label{tmp6},
\end{align}
where the factors $\chi_i(\beta,\varphi_i)$ become independent of particle size and electron velocity in the limit of very small particles. 

To obtain the scaling properties of EELS and CL probabilities for 2D structures, we first notice that, because the permittivity depends on the sample thickness $d$, a transformation of the structure size must reflect on the screened interaction properties. Indeed, by using the scaling properties of the eigenpotentials and the fact that $\eta_i$ does not depend on particle size, we obtain the relation
\begin{align}
\tilde{\mathcal{W}}(\rb,\rb',\omega; \sigma)=\frac{1}{\mu}\mathcal{W}(\rb/\mu,\rb'/\mu,\omega; \sigma/\mu)\label{2dWscaling}
\end{align}
between the screened interaction of a system with size $\mu D$ ($\tilde{\mathcal{W}}$ on the left-hand side) and the one for a system with size $D$ ($\mathcal{W}$ on the right-hand side), where we have explicitly indicated the dependence on the surface conductivity $\sigma$ of the 2D material. Equation~\ref{2dWscaling} closely resembles its 3D equivalent in Eq.~(\ref{wscal}), except for the difference that the equality in Eq.~(\ref{2dWscaling}) only holds if all lengths and the conductivity are simultaneously scaled. This leads to the EELS and CL scaling relations (also indicating an explicit dependence on $\sigma$)
\begin{align}
\tilde{\Gamma}_{\rm EELS} (\Rb_0,v,\omega;\sigma)&=\frac{1}{\mu}\Gamma_{\rm EELS} (\Rb_0/\mu ,v/\mu ,\omega; \sigma/\mu),\label{eels2D}\\
\frac{d\tilde{\Gamma}_{\rm CL}(\Rb_0,v,\omega;\sigma)}{d\omega }&=\sum_{n=0}^\infty \mu^{n+2}\,\frac{d\Gamma_{{\rm CL}}^{(n)}(\Rb_0/\mu,v/\mu,\omega;\sigma/\mu)}{d\omega}, \label{ffarl2D}
\end{align}
which show the same behavior with $\mu$ as in Eqs.~(\ref{quasieelsscal}) and (\ref{ffarl}), respectively, because of the combined change in the dimensionality of the sample and the dependence of EELS and CL probabilities with conductivity (see Eqs.~(\ref{eelsquasi}) and (\ref{ffquasi})).

\section{Mie Modes in Dielectric Spheres}
\renewcommand{\theequation}{G\arabic{equation}}
\label{AppendixG}

For a self-standing sphere of diameter $D$ and permittivity $\epsilon$, Mie theory \cite{M1908} gives the mode field distributions. Following a standard procedure \cite{B12_2}, one identifies electric and magnetic modes indexed by the orbital and azimuthal numbers $\ell$ and $m$ at $m$-independent frequencies determined by the conditions \cite{J99}
\begin{subequations}
\label{wem}
\begin{align}
&h_\ell^{(1)}(x_i)\partial_{x_i}\big[x_ij_\ell(\sqrt{\epsilon}x_i)\big]-\epsilon j_\ell(\sqrt{\epsilon}x_i)\partial_{x_i}\big[x_i h_\ell^{(1)}(x_i)\big]=0,\label{we} \\
&h_\ell^{(1)}(x_i)\partial_{x_i}j_\ell(\sqrt{\epsilon}x_i)-j_\ell(\sqrt{\epsilon}x_i)\partial_{x_i} h_\ell^{(1)}(x_i)=0,\label{wm}
\end{align}
\end{subequations}
where $x_i=\omega_iD/2c$, $\omega_i$ are mode frequencies labelled by $i$, while $j_\ell$ and $h_\ell^{(1)}$ are spherical Bessel and Hankel functions. The solutions of Eqs.~(\ref{wem}) are complex eigenfrequencies in general, which can be dealt with through the quasinormal-mode formalism \cite{CLM98,GH15,LYV18,FHD19}. In the present work, we focus on modes with a low level of radiation losses and consider only the real part of $\omega_i$ (see supplementary Fig.~\ref{FigS2}). By approximating the spherical Bessel and Hankel functions in Eqs.~(\ref{wem}) for small arguments \cite{AS1972}, the electric mode frequencies are found to satisfy the equation
\begin{align}
x_i^2\bigg[\epsilon^2 \frac{\ell}{2\ell+3}+ \epsilon\frac{3(2\ell+1)}{(2\ell+3)(2\ell-1)}-\frac{\ell+1}{2\ell-1}\bigg]=2(\epsilon\ell+\ell+1),\nonumber
\end{align}
which leads to $\omega_i \approx (c/2D\sqrt{\epsilon})\sqrt{2(2\ell+3)}$ for $\epsilon\gg 1$.

\section{Electromagnetic Green Tensor in the Quasistatic Limit}
\renewcommand{\theequation}{H\arabic{equation}}
\label{AppendixH}

The Green tensor defined in Eq.~(\ref{gt}) permits calculating the electric field $\Eb(\rb,\omega)=-4\pi\ii\omega\int d^3\rb'\,G(\rb,\rb',\omega)\cdot\jb(\rb,\omega)$ produced by a current density $\jb(\rb,\omega)$. In the electrostatic limit, the screened interaction $\mathcal{W}(\rb,\rb',\omega)$ relates the induced potential $\phi(\rb,\omega)=\int d^3\rb'\,\mathcal{W}(\rb,\rb',\omega)\rho(\rb',\omega)$ to the generating charge density $\rho(\rb',\omega)$. Expressing the latter as $\rho(\rb',\omega)=(1/\ii\omega)\,\nabla'\cdot\jb(\rb',\omega)$ in virtue of the continuity equation, integrating over $\rb'$ by parts, and calculating the electric field as $\Eb(\rb,\omega)=-\nabla\phi(\rb,\omega)$, we find $\Eb(\rb,\omega)=-4\pi\ii\omega\int d^3\rb'\,G(\rb,\rb',\omega)\cdot\jb(\rb,\omega)$ with $G(\rb,\rb',\omega)= \nabla_\rb \otimes \nabla_{\rb'} \mathcal{W}(\rb,\rb',\omega)/4\pi\omega^2$, which can therefore be understood as the electrostatic approximation to the Green tensor. Combined with Eq.~(\ref{wscal}), we readily obtain the scaling $\tilde{G}(\rb,\rb',\omega)=\mu^{-3}\,G(\rb/\mu,\rb'/\mu,\omega)$ in this limit.

\section{Free-Electron Coupling to an Optical Mode}
\renewcommand{\theequation}{I\arabic{equation}}
\label{AppendixI}

We provide a more general derivation of the excitation probability, leading to the same result as obtained above for dielectric cavities (i.e., Eq.~(\ref{Peelscldiel})). Consider an optical mode $i$ of frequency $\omega_i$ characterized by an electric field $\Eb_i(\rb)$ that satisfies the normalization condition $\int d^3\rb\,\epsilon(\rb)|\Eb_i(\rb)|^2=1$. The coupling to a swift electron that moves with constant velocity $\vb\parallel\zz$ is then described through the interaction Hamiltonian \cite{paper339} $\hat{\mathcal{H}}'(t)=-(\ii e/\omega_i)\,\vb\cdot\big[\Eb_i(\rb)\hat{a}-\Eb_i^*(\rb)\hat{a}^\dagger\big]$, where the position vector $\rb$ needs to be evaluated at the time-dependent electron position $\Rb_0+\vb t$. Starting from the mode in the ground state $\ket{0}$, the post-interaction amplitude of the $\ket{1}$ Fock state is given by $\alpha_i=-\ii\hbar^{-1}\int_{-\infty}^\infty dt\,\bra{1}\hat{\mathcal{H}}'(t)\ket{0}\ee^{-\ii\omega_it}$. Combining these elements, the excitation probability reduces to $P_i\equiv|\alpha_i|^2=\hbar^{-1}(ev/\omega_i)^2\big|\int_{-\infty}^\infty dt\,E_{i,z}(\Rb_0,vt)\ee^{-\ii\omega_it}\big|^2$, which can readily be recast into Eq.~(\ref{Peelscldiel}) with $F_i(\varphi_i)$ and $\varphi_i$ defined by Eqs.~(\ref{Fivarphii}) by expressing $\Eb_i(\rb)=\tilde\Eb_i(\rb/D)/D^{3/2}$ in terms of the dimensionless scaled field $\tilde\Eb_i(\rb/D)$.

\section{EELS Probability in a Metal-Coated Cylindrical Cavity}
\renewcommand{\theequation}{J\arabic{equation}}
\label{AppendixJ}

We investigate the excitation probability of the electromagnetic modes supported by a dielectric cylinder (real permittivity $\epsilon$, radius $a$, length $D$) coated by a perfect electric conductor and traversed by an electron moving parallel to the axis at a distance $R_0$. The cavity supports TE and TM modes, but only the latter have a nonvanishing electric field along the axis and can therefore couple to the electron. Adopting well-known expressions for the electric field distribution of cylindrical TM waves \cite{paper047} and using cylindrical coordinates $\rb=(R,\varphi,z)$ (with the cavity defined by $0<z<D$ and $R<a$), we impose the condition of vanishing surface-parallel electric field components at the boundaries and find the cavity-mode fields
\begin{align}
\Eb^{\rm TM}_{nmk}(\rb) =  \frac{1}{\sqrt{\pi\epsilon I_{nmk}D}} \Big[ &nqJ'_{m}(Q_{mk} R)  \sin(nqz) \hat{\Rb}+\ii \frac{n q}{\qmk}\frac{m}{R}J_m(\qmk R)\sin\left( nqz\right) \boldsymbol{\hat{\varphi}}  \nonumber\\
&-\qmk J_m(\qmk R )\cos(nqz) \zz \vphantom{\frac{1}{2}}\Big] \;\ee^{\ii m \varphi}, \label{cyltm2}
\end{align}
where $q=\pi/D$, $n$ indicates the number of nodes in the $z$ component, $m$ is the azimuthal number, $k$ labels different radial modes of transverse wave vectors $Q_{mk}=z_{mk}/a$ determined by the condition $J_m(z_{mk})=0$, $J_m$ are Bessel functions, and
\begin{align}
I_{nmk}
&=\frac{1}{2}\Big[\Big(\frac{n\pi a}{D}\Big)^2+(1+\delta_{n0})\,z_{mk}^2\Big]\,J_{m-1}^2(z_{mk})\label{normintegral}
\end{align}
is a normalization constant ensuring the condition $\epsilon\int_V d^3\rb\,\big|\Eb^{\rm TM}_{nmk}(\rb)\big|^2=1$, with the integral running over the cavity volume $V$. The mode frequencies $\omega_{nmk}=c\sqrt{(Q_{mk}^2 + n^2 q^2)/\epsilon}$ are determined by imposing the conservation of the total electromagnetic wave vector. As shown above, the excitation probability is given by Eq.~(\ref{Peelscldiel}) with the $F_i(\varphi_i)$ factor specialized to the $i=\{nmk\}$ mode. By taking the Fourier transform of Eq.~(\ref{cyltm2}), the dimensionless function in Eq.~(\ref{Fi}) becomes
\begin{align}
F_{nmk}(\varphi_{nmk})=\frac{2z_{mk}^2D^2}{\pi\epsilon I_{nmk}a^2} J^2_m\left(\frac{z_{mk} R_0}{a}\right)\varphi_{nmk}\frac{1+(-1)^{n+1}\cos(\varphi_{nmk})}{[n^2\pi^2-\varphi_{nmk}^2]^2},\label{FCCnmk}
\end{align}
where $\varphi_{nmk}=\omega_{nmk}D/v$. 

\section{Excitation Probability for a Free-Electron-Induced s-p Transition}
\renewcommand{\theequation}{K\arabic{equation}}
\label{AppendixK}

We consider a model system comprising ground and excited states described by one-electron hydrogenic s and p orbitals $\psi_s(\rb)=\Cs\ee^{- r/D}$ and $\psi_p(\rb)=\Cp\ee^{-r/2D}\rb\cdot\nn$, where $\Cs=(\pi D^3)^{-1/2}$ and $\Cp=(32\pi D^5 )^{-1/2}$ are normalization constants, while $\nn$ is a unit vector that defines the direction of the transition dipole. To satisfy the Schr\"odinger equation, the transition energy $\hbar\omega_{sp}$ must satisfy the identity $\omega_{ sp }\langle\psi_p|\rb| \psi_s\rangle=-\hbar\langle \psi_p|\nabla_\rb| \psi_s\rangle/\me $, which, for the orbitals under consideration, leads to $\omega_{sp}= 3\hbar/(8\me D^2)$. We describe the s-p transition through the minimal coupling interaction Hamiltonian $\hat{\mathcal{H}}'(\rb,t)=-(\ii\hbar e/2\me c)(\nabla\cdot\Ab(\rb,t)+\Ab(\rb,t)\cdot\nabla)-e\phi(t)$ in which we neglect $A^2$ terms, while $\Ab$ and $\phi$ are the vector and scalar potentials of the electromagnetic field produced by the electron, treated as a point charge that moves with constant velocity $v$ along $z$ (nonrecoil approximation \cite{paper149}). 

We take the system to be initially prepared in the s state and calculate the amplitude of the p state $\alpha_p(t)$ within first-order perturbation theory. The post-interaction amplitude then reads $\alpha_p(\infty)\approx-\ii\hbar^{-1}\int_{-\infty}^\infty dt\,\int d^3\rb\,\psi_p^*(\rb)\hat{\mathcal{H}}'(\rb,t)\psi_s(\rb)\ee^{\ii\omega_{sp}t}$, which can be rewritten in terms of the electric field $\Eb_{sp}(\rb)=-4\pi \ii \omega_{sp}\int d^3\rb' G_0(\rb,\rb',\omega_{sp})\jb_{sp}(\rb')$ produced by the transition current
$\jb_{sp}(\rb)=(\ii \hbar e/2\me)\left[\psi_s\nabla \psi_p-\psi_p\nabla \psi_s\right]$ as
\begin{align}
\alpha_{p}(\infty)=\frac{e}{\hbar \omega_{sp}}\int_{-\infty}^\infty dz\,E^*_{sp,z}(\Rb_0,z)\,\ee^{\ii \omega_{sp}z/v}.\label{alphp}
\end{align}
To derive this result, we have considered the vector potential
\[\Ab(\rb,t)=2 ec\int_{-\infty}^\infty d\omega \int_{-\infty}^\infty dz' \ee^{\ii\omega (z'/v-t)} G_{0}(\rb,\Rb_0,z',\omega)\]
produced by an electron traveling in vacuum and crossing the transverse position $\Rb_0$ at $t=0$, and we have used the Onsager reciprocity relation $G_{0,ii'}(\rb,\rb')=G_{0,i'i}(\rb',\rb)$. We note that an incomplete calculation would have led to a different definition of the transition field containing $G^*_0(\rb,\rb',\omega_{sp})$ instead of $G_0(\rb,\rb',\omega_{sp})$. However, the two expressions differ by a term $-4\pi\ii \omega_{sp}\int d^3\rb'\; {\rm Im}\{G_0(\rb,\rb',\omega_{sp})\}\,\jb_{sp}(\rb')$ that contains only real transverse photons with a dispersion $\omega=kc$, so they do not couple to the free electron. Reassuringly, a coupling coefficient of the form in Eq.~(\ref{alphp}) corresponds to the one adopted in previous works\cite{paper339} to describe the coupling of free electrons to atomic transitions in a quantum-optics framework. 

We can now compute the transition electric field from the atomic orbitals combined with the electromagnetic Green function in the momentum representation. We find the result $\alpha_p(\infty)=\boldsymbol{\alpha}\cdot\nn$ with
\begin{align}
\boldsymbol{\alpha}=-\frac{8e^2}{v}\Cs\Cp\int & d^2\Qb \frac{\ee^{-\ii\Qb\cdot\Rb_0}}{Q^2+(\omega_{sp}/v\gamma)^2}\\ 
&\times\left[ \frac{ v}{D\me c^2}\frac{1}{(g^2 + q^2)^2}\left(\zz - \qb \frac{4\omega_{sp}/v}{g^2 + q^2}\right)+ \qb\left(\frac{\omega_{sp}}{2 \me c^2}- \frac{1}{\hbar}\right)\frac{4g }{(g^2 + q^2)^3}\right] , \nonumber
\end{align}
where $\qb = (\Qb,\omega_{sp}/v)$. The azimuthal part of the $\Qb$ integral can be expressed in terms of Bessel functions, so the excitation probability $P_{sp}=|\alpha_p(\infty)|^2$ finally reduces to Eq.~(\ref{Psp}), where
\begin{align}
F_{sp}(\varphi_{sp})&=\frac{288}{\varphi_{sp}}\Bigg| \int_0^\infty ds \,\frac{s}{[s^2+(\varphi_{sp}/\gamma)^2][9/4+s^2+\varphi_{sp}^2]} \label{Fsptmp1}\\
&\!\!\!\!\!\!\!\!\!\!\times \Bigg\{\ii s \varphi_{sp}(\nn \cdot \hat{\Rb}_0)J_1\left(\frac{sR_0}{D}\right)-\varphi^2_{sp}(\nn\cdot \zz)J_0\left(\frac{sR_0}{D}\right)\bigg[\frac{1}{\gamma^2}+\frac{\hbar \omega_{sp}}{6\me c^2}-\frac{4\beta^2 }{9}\big(s^2+\varphi_{sp}^2\big)\bigg]\Bigg\}\Bigg|^2 \nonumber
\end{align}
and $\varphi_{sp}=\omega_{sp}D/v$. We note that the result in Eq.~(\ref{Fsptmp1}) can be alternatively expressed as the squared Fourier transform of the scaled field $\tilde{\Eb}_{sp}(\rb/D)=(D^2/e)\Eb_{sp}(\rb)$; namely, $F_{sp}(\varphi_{sp})=\varphi_{sp}^{-1}|\int_{-\infty}^\infty du_z \tilde{\Eb}_{sp}(\Rb/D,u_z)\,\ee^{-\ii\omega_{sp} u_z /v}|^2$, which becomes scale invariant only in the $c\rightarrow\infty$ limit.

In the long-distance limit ($R_0\gg D$), we have $\varphi_{sp}\ll1$, and only the $s\ll1$ region contributes to the integral, which can then be approximated to yield
\begin{align}
P_{sp}(\Rb_0,v) \approx \bigg(\frac{2\omega_{sp}}{\hbar v^2\gamma}\bigg)^2 \bigg[(\zz \cdot \db)^2 \frac{1}{\gamma^2} K_0^2\Big(\frac{\omega_{sp} R_0}{v\gamma }\Big)+(\Rb_0\cdot\db)^2 K_1^2\Big(\frac{\omega_{sp} R_0}{v\gamma }\Big)\bigg],\label{Pspdipn}
\end{align}
where $\db=-e\int d^3\rb\,\rb\,\psi_p^*(\rb)\psi_s(\rb)=-(2^{15/2}/3^5)\,eD\,\nn$ is the transition dipole and $K_m$ are modified Bessel functions. Equation~\ref{Pspdipn} coincides with the frequency-integrated EELS probability in a dipolar particle described by a polarizability $\alpha(\omega) = (\db\otimes\db/\hbar)[1/(\omega_{sp} - \omega + \ii 0^+)+1/(\omega_{sp} + \omega + \ii 0^+)]$ (i.e., hosting a single resonance at the same frequency and with the same transition dipole).

\section{Free-Electron-Induced Excitation Probability of Waveguide Modes}
\renewcommand{\theequation}{L\arabic{equation}}
\label{AppendixL}

As a specific instance of 3D dielectric structure, we consider a 1D waveguide running along $z$. We focus on a band in which waveguide modes are labeled by the parallel wave vector $\kpar$. Mode fields can be written as $\Eb_{\kpar}(\Rb)\ee^{\ii\kpar z}/\sqrt{L}$ (i.e., with $i\to\kpar$), where $L$ is the quantization length along $z$. For dielectric waveguides, the condition $\int d^2\Rb\,\epsilon(\Rb)|\Eb_{\kpar}(\Rb)|=1$ is imposed by mode normalization, whereas in waveguides with a dispersive permittivity that condition can be more involved \cite{BS06_2}. Using this field in Eq.~(\ref{Fi}), integrating over $z$ along an effective interaction length $L_{\rm eff}$, inserting the result into Eq.~(\ref{Peelscldiel}), and summing over $\kpar$ modes by adopting the prescription $\sum_{\kpar}\to(L/2\pi)\int d\kpar$, we find
\begin{align}
P_{{\rm EELS}}=4\alpha c\int_{-\infty}^\infty\frac{d\kpar}{\omega_{\kpar}}|E_{\kpar,z}(\Rb_0)|^2\,\frac{\sin^2(\Delta L_{\rm eff}/2)}{\Delta^2},
\label{lastlast}
\end{align}
where $\Delta=\kpar-\omega_{\kpar}/v$, the lateral position of the electron trajectory is defined by $\Rb_0$, and $\omega=\omega_{\kpar}$ gives the dispersion relation of the waveguide band. The quantization length $L$ has disappeared from this result and only $L_{\rm eff}$ remains. We now approximate $\sin^2(\Delta L_{\rm eff}/2)/\Delta^2\approx2\pi L_{\rm eff}\delta(\Delta)$ assuming $\kpar L_{\rm eff}\gg1$. Phase-matching with the electron field occurs under the condition $\Delta=0$ (i.e., the electron line $\omega_{\kpar}=\kpar v$), which defines a crossing point with $\kpar=k_c$ defined by the geometrical construction in Fig.~\ref{Fig6}a. Finally, we divide the probability by $L_{\rm eff}$ and directly obtain the excitation probability per unit of electron path length in Eq.~(\ref{Pwaveguide}).

When the electron line is tangent to the dispersion relation with a general behavior near $\kpar=k_c$ given by $\omega_{\kpar}=\kpar v+\zeta(\kpar-k_c)^n$ with $n=1,2,\cdots$, changing the integration variable in Eq.~(\ref{lastlast}) to $\theta=(\kpar-k_c)^n L_{\rm eff}\zeta/2v$ yields Eq.~(\ref{lastlastlast}) with a scaling $P_{{\rm EELS}}\propto L_{\rm eff}^{2-1/n}$.

\end{widetext}

\section*{Acknowledgments}
This work has been supported in part by ERC (Advanced Grants 789104-eNANO and 101019932-QEWS), the European Commission (project No. 101017720-EBEAM), the Spanish MICINN (PID2020-112625 GB-I00 and Severo Ochoa CEX2019-000910-S), the Dutch Research Council (NWO), the Catalan CERCA Program, and Fundaci\'os Cellex and Mir-Puig.


%

\begin{figure}[ht!]
\centering
\includegraphics[width=0.79\textwidth]{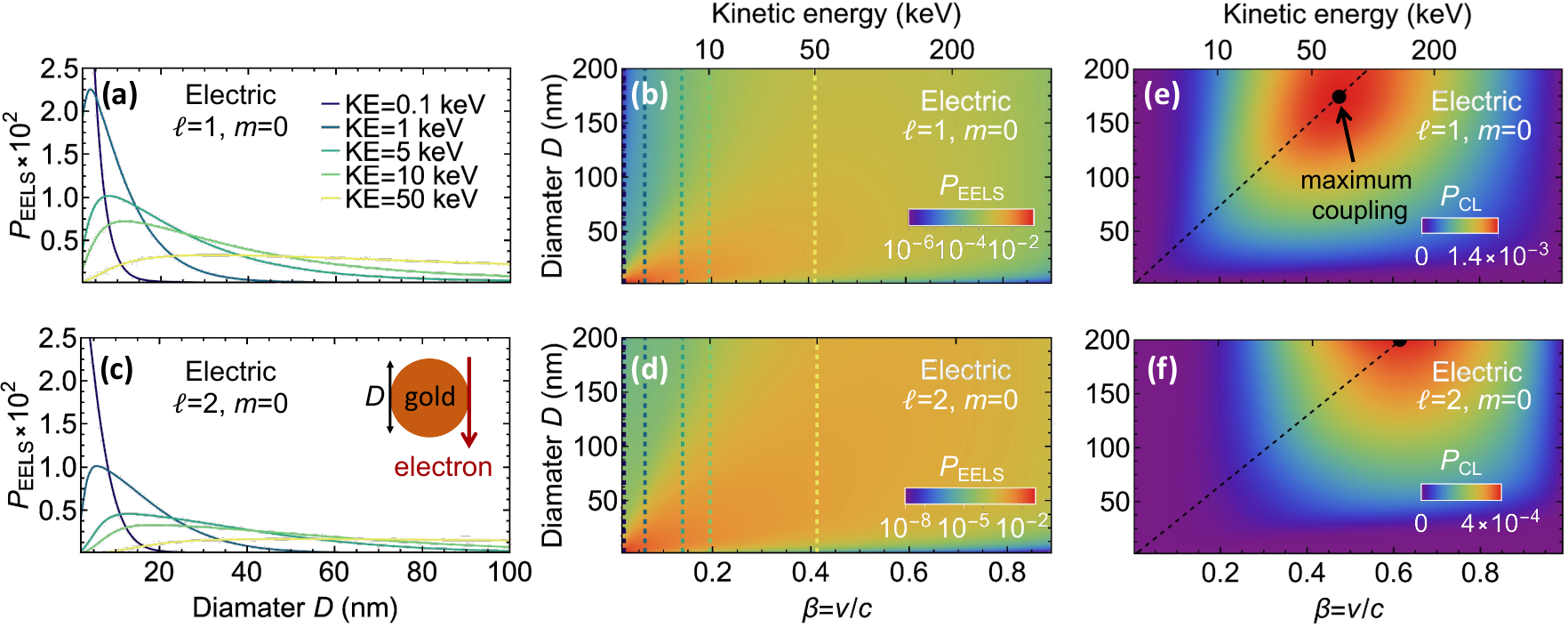}
\caption{{\bf EELS and CL in a gold sphere.} {\bf (a-d)}~Frequency-integrated EELS probability for an electron passing grazingly to a gold nanosphere (see inset in panel (c)) and coupling either to the electric dipolar ($\ell=1$) $m=0$ mode (a,b) or to the electric quadrupolar ($\ell=2$) $m=0$ mode (c,d) as a function of sphere diameter $D$ and electron kinetic energy (KE). Panels (a,c) show cuts along the color-matched dashed lines in (b,d). {\bf (e,f)}~CL emission probabilities for the modes discussed in (b,d). Black dots indicate the positions at which a maximum CL coupling is reached, while black-dashed lines connect the origin of axes with such points. Probabilities are computed from closed-form analytical expressions \cite{paper021} using tabulated experimental data for the dielectric function of gold \cite{JC1972}.}
\label{FigS1}
\end{figure}

\begin{figure}[ht!]
\centering
\includegraphics[width=0.65\textwidth]{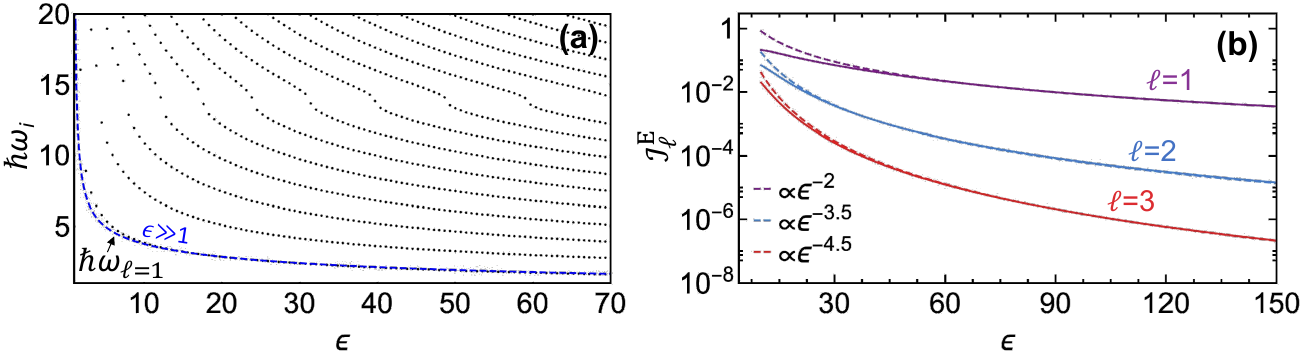}
\caption{{\bf Energies and Mie scattering coefficients for electric modes.} {\bf (a)}~Mode energies $\hbar\omega_i$ (black dots) satisfying the resonance condition ${\rm Re}\{[h_\ell^{(+)}(x_i)\partial_{x_i}[x_ij_\ell(\sqrt{\epsilon}x_i)]]/[\epsilon j_\ell(\sqrt{\epsilon}x_i)\partial_{x_i}[x_i h_\ell^{(+)}(x_i)]]\}=1$ with $x_i=\omega_iD/2c$ (see Ref.~\citenum{paper021}) for a dielectric sphere of diameter $D=128$~nm as a function of its permittivity $\epsilon$. We also plot the approximate solution $x_i^2\left\{(\ell+1)/(2\ell-1)-3\epsilon (2\ell+1)/[(2\ell+3)(2\ell-1)]-\epsilon^2 \ell/(2\ell+3)\right\}+4.5(\epsilon\ell+\ell+1)\approx 0$ for $\ell=1$ (dashed blue curve). {\bf (b)}~Dependence of the integral $\mathcal{I}_\ell^{\rm E}=\int_{\omega_i-\delta}^{\omega_i+\delta} d\omega \, t_\ell^{\rm E}(\omega)/\omega_i$ on $\epsilon$. We take $\hbar\delta=0.5$~eV and $\omega_i$ from the blue curve in (a).}
\label{FigS2}
\end{figure}

\begin{figure}[ht!]
\centering
\includegraphics[width=1.00\textwidth]{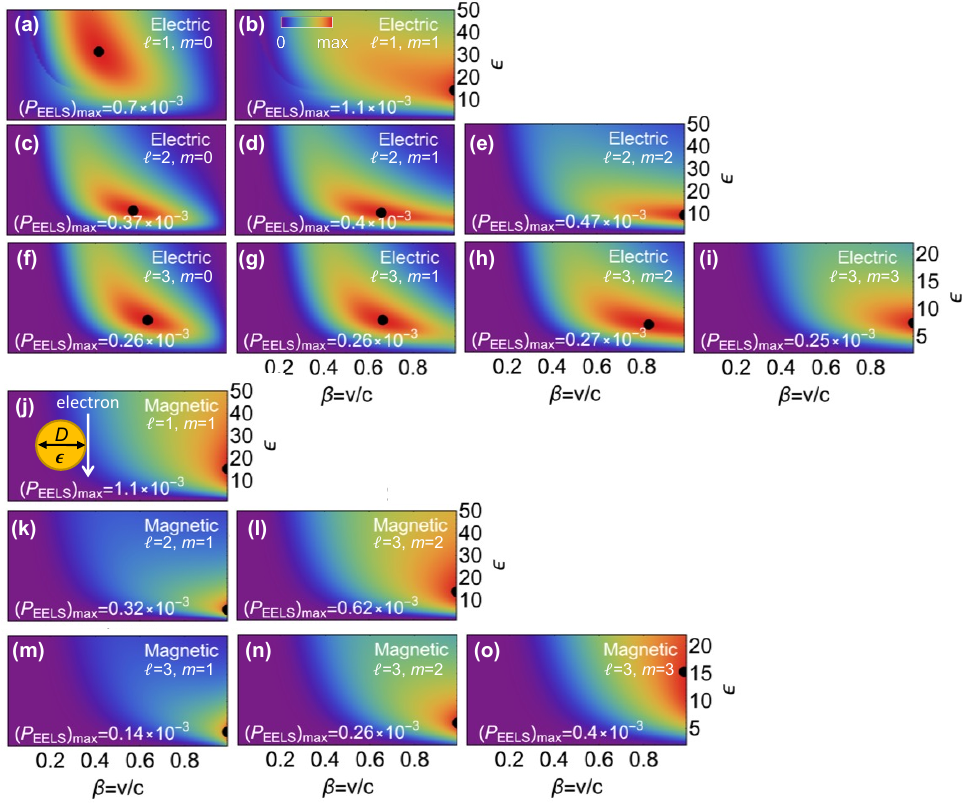}
\caption{{\bf  Electron coupling to a dielectric sphere.} Probability that an electron passing grazingly to a self-standing dielectric sphere of diameter $D=128$~nm excites a Mie mode as a function of the (real) permittivity $\epsilon$ and the scaled electron velocity $\beta=v/c$ (see sketch in (j)). We present results for different electric {\bf (a-i)} and magnetic {\bf (j-o)} modes. The coupling probability is computed by integrating each EELS spectrum over the $\hbar\omega_1\pm 0.5$~eV range, where $\omega_1$ refers to the lowest-frequency mode in Fig.~\ref{FigS2}a.}
\label{FigS3}
\end{figure}

\begin{figure}[ht!]
\centering
\includegraphics[width=0.8\textwidth]{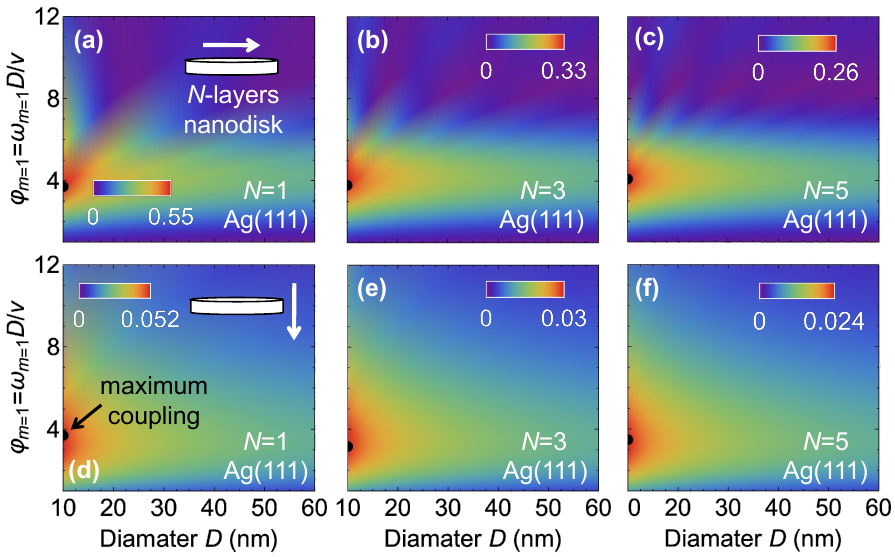}
\caption{{\bf Electron coupling to thin silver nanodisks.} We calculate the probability for a free electron to couple to the $m=1$ mode supported by a nanodisk as a function of disk diameter $D$ and the phase $\varphi_{m=1}=\omega_{m=1}D/v$ (see Eq.~( xxx12 in the main text). The probability is presented for parallel {\bf (a-c)} and perpendicular {\bf (d-e)} trajectories as defined in Fig.~4 with disks consisting of $N=1$ (a,d) $N=3$ (b,e), and $N=5$ (c,f) atomic layers. The optical response of each disk is modeled through the two-dimensional conductivity $\sigma(\omega)=(\ii e^2/\hbar )\,\omega_D/(\omega-\omega_g-\ii \gamma)$ using the values reported in Table 1 for $\omega_{\rm D}\propto N$, $\omega_g$, and $\gamma$.}
\label{FigS4}
\end{figure}

\begin{figure}[ht!]
\centering
\includegraphics[width=0.8\textwidth]{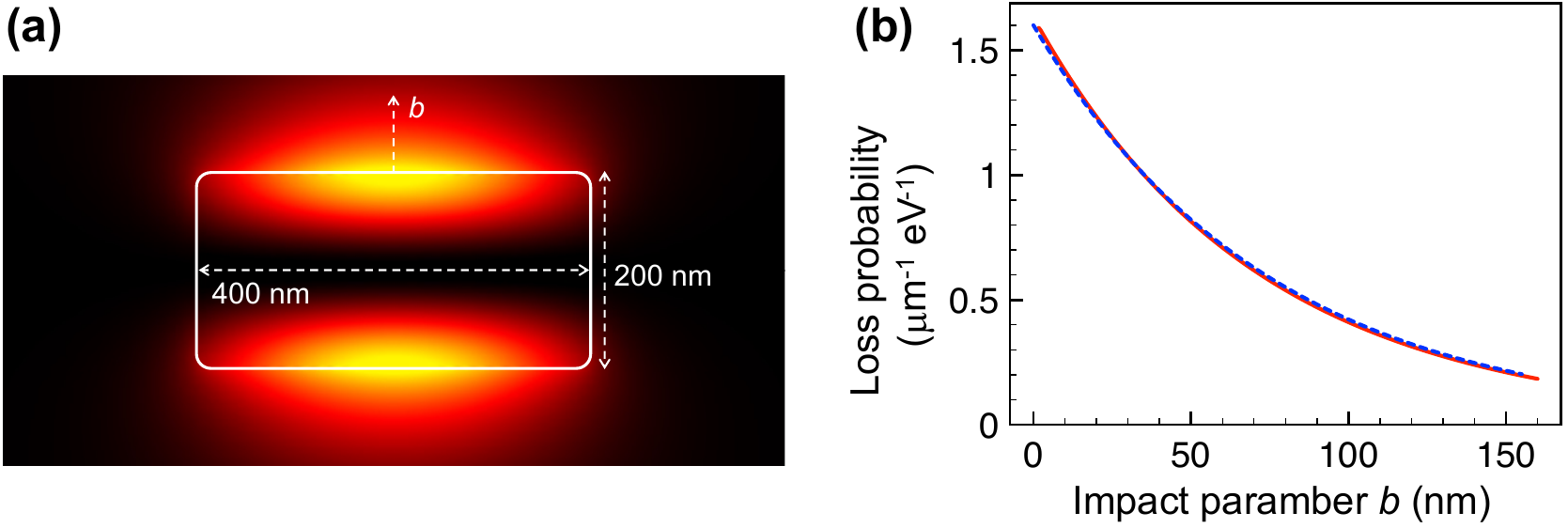}
\caption{{\bf Mode profile in a silicon waveguide.} We consider the same Si waveguide as in Fig.~6b in the main text, where a small imaginary part is added to the dielectric function ($\epsilon=12+0.1\,\ii$) to better visualize the excitation features. Here, we investigate the spatial characteristics of the mode corresponding to the lowest-energy feature (photon energy $\hbar\omega\approx0.97$~eV). {\bf (a)} Spatial distribution of the loss probability. {\bf (b)} Impact-parameter dependence of the loss probability (red curve) for electron-beam positions along the upper dashed arrow in (a). We also plot the exponential $\propto\ee^{-2b/\lambda_\perp}$ (dashed blue curve) with $\lambda_\perp=150$~nm.}
\label{FigS5}
\end{figure}

\end{document}